\documentclass[useAMS,usenatbib]{mn2e}
\usepackage{epsfig,lscape,amsmath,amssymb}
\usepackage{graphicx,longtable,times,threeparttable,multirow}
\usepackage{rotating,color}
\usepackage[T1]{fontenc}
\usepackage{aecompl}
\usepackage{bm}

\DeclareGraphicsExtensions{.ps, .eps,.eps.gz,.epsi}

\begin{document}

\title[]{Estimating stellar atmospheric parameters, absolute magnitudes and elemental abundances 
from the LAMOST spectra with Kernel-based Principal Component Analysis}
\author[Xiang et al.]{M.-S. Xiang$^{1}$\thanks{LAMOST Fellow}\thanks{E-mail: msxiang@nao.cas.cn}, X.-W. Liu$^{2,3}$, 
J.-R. Shi$^{1}$, H.-B. Yuan$^{4}$, Y. Huang$^{2}$,  A.-L. Luo$^{1}$, 
\newauthor H.-W. Zhang$^{2}$, Y.-H. Zhao$^{1}$, J.-N. Zhang$^{1}$, 
J.-J., Ren$^{2}$\footnotemark[1], B.-Q. Chen$^{2}$\footnotemark[1], C. Wang$^{2}$, 
 \newauthor J. Li$^{5}$, Z.-Y. Huo$^{1}$, W. Zhang$^{1}$, J.-L. Wang$^{1}$, 
Y. Zhang$^{6}$, Y.-H. Hou$^{6}$, Y.-F. Wang$^{6}$        
\\ \\
$1$ National Astronomical Observatories, Chinese Academy of Sciences, Beijing 100012, P. R. China \\
 $2$ Department of Astronomy, Peking University, Beijing 100871, P. R. China \\
$3$ Kavli Institute for Astronomy and Astrophysics, Peking University, Beijing 100871, P. R. China \\
$4$ Department of Astronomy, Beijing Normal University, Beijing 100875, P. R. China \\
$5$ Department of Space Science and Astronomy, Hebei Normal University, Shijiazhuang 050024, P. R. China \\
$6$ Nanjing Institute of Astronomical Optics \& Technology, National Astronomical Observatories,
    Chinese Academy of Sciences, Nanjing 210042, P. R. China \\}
\date{Received:}

\pagerange{\pageref{firstpage}--\pageref{lastpage}} \pubyear{2015}

\maketitle

\label{firstpage}

\begin{abstract}{
Accurate determination of stellar atmospheric parameters and elemental abundances is 
crucial for Galactic archeology via large-scale spectroscopic surveys. In this paper, 
we estimate stellar atmospheric parameters --- effective temperature $T_{\rm eff}$,
surface gravity log\,$g$ and metallicity [Fe/H], absolute magnitudes 
${\rm M}_V$ and ${\rm M}_{K{\rm s}}$, $\alpha$-element to metal (and iron) abundance ratio [$\alpha$/M] (and [$\alpha$/Fe]), 
as well as carbon and nitrogen abundances [C/H] and [N/H] from the LAMOST spectra with 
a multivariate regression method based on kernel-based principal component analysis, 
using stars in common with other surveys (Hipparcos, $Kepler$, APOGEE) as training data sets. 
Both internal and external examinations indicate that given a spectral signal-to-noise 
ratio (SNR) better than 50, our method is capable of delivering stellar 
parameters with a precision of $\sim$100\,K for $T_{\rm eff}$, $\sim$0.1\,dex for 
log\,$g$, 0.3 -- 0.4\,mag for ${\rm M}_V$ and ${\rm M}_{K{\rm s}}$, 
0.1\,dex for [Fe/H], [C/H] and [N/H], and better than 0.05\,dex 
for [$\alpha$/M] ([$\alpha$/Fe]). The results are satisfactory even for a spectral SNR of 20. 
The work presents first determinations of [C/H] and [N/H] abundances 
from a vast data set of LAMOST, and, to our knowledge, the first reported implementation of absolute 
magnitude estimation directly based on the observed spectra. The derived stellar parameters for millions of 
stars from the LAMOST surveys will be publicly available in the form of value-added catalogues.}
\end{abstract}

\section{Introduction}

\label{sect:intro}
Accurate estimation of stellar atmospheric parameters and elemental abundances from large samples of high-to-low resolution 
spectra collected by modern large spectroscopic surveys, such as the LAMOST Experiment for Galactic Understanding 
and Exploration \citep[LEGUE;][]{Deng+2012, Zhao+2012}, the Sloan Extension for Galactic Understanding and Exploration 
\citep[SEGUE;][]{Yanny+2009}, the Apache Point Observatory Galactic Evolution Experiment \citep[APOGEE][]{Majewski+2010}, 
the Radial Velocity Experiment \citep[RAVE;][]{Steinmetz+2006}, 
the High Efficiency and Resolution Multi-Element Spectrograph survey \citep[HERMES;][]{Zucker2012,Freeman2012}, 
and the Gaia Radial Velocity Spectrograph (RVS) survey \citep[e.g.][]{Bailer-Jones+2013}, is of vital importance for Galactic archaeology. 
For this purpose, various methods have been developed, based on either a $\chi^2$-related algorithm 
\citep{Allende_Prieto+2006, Lee+2008a, Zwitter+2008, Wu+2011, Wu+2014, Xiang+2015a, Garcia_Perez+2015} 
or a multivariable analysis (e.g. principal component analysis, PCA) incorporated into 
a neural network scheme \citep{Recio-Blanco+2006, Re_Fiorentin+2007, Manteiga+2010, 
Yang_Li+2015, Lu_Li+2015, Li_Lu+2015, Bu+2015, Liuchao+2015, Recio-Blanco+2015}.   

In spite of the efforts, in the case of low-resolution spectra, the accuracy remains to be improved, 
especially for parameter such as the surface gravity log\,$g$ that has a relatively weak 
dependence on the observed spectral features. 
Typical uncertainties of log\,$g$ estimates yielded by most pipelines are about 0.2\,dex or larger \citep{Lee+2008a,Lee+2008b, 
Zwitter+2008, Kordopatis+2013, Xiang+2015a, Gao+2015}. 
Patterns of systematic errors in $\log\,g$ are clearly visible in the results of SEGUE Stellar Parameter Pipeline (SSPP) 
for the SDSS SEGUE spectra and those of LAMOST Stellar Parameter Pipeline at PKU (LSP3) for the LAMOST 
spectra \citep{Xiang+2015a, Ren+2016}. 
Those large errors induce substantial uncertainties in other inferred parameters, 
such as stellar distance and age. The errors could also cause  
problems for classifying stars. For example, given the relatively narrow ($\sim$0.5\,dex) range of values of log\,$g$ of 
main-sequence turn-off/subgiant stars, samples of such stars selected based on stellar atmospheric parameters 
yielded by the aforementioned pipelines may suffer from significant contaminations from the numerous 
long-lived main-sequence dwarfs \citep{Xiang+2015c}.  
Similarly, for the evolved stars, as a result of the uncertainties in the derived parameters, stars of the red giant branch (RGB), 
the red clump (RC) and of the asymptotic giant branch (AGB) are mixed together 
in the $T_{\rm eff}$ -- $\log\,g$ diagram 
and difficult to disentangled from each other \citep{Huang+2015a}. 
Finally, little efforts has been made hitherto to estimate stellar luminosity directly 
from the stellar spectra, or to estimate the individual elemental abundances from low-resolution spectra.

One of the error sources for stellar atmospheric parameter determination comes from uncertainties 
in the template/training sets of spectra adopted. 
Both synthetic and empirical spectra have been widely used as the spectral templates or training sets. 
For synthetic templates, the fidelity of a spectrum for a given set of atmospheric parameters ($T_{\rm eff}$, log\,$g$, [Fe/H], [$\alpha$/Fe]) 
is limited by our knowledge of the often complicated stellar astrophysics and atomic/molecular opacities. 
A comprehensive comparison with the observed spectra of stars with accurate, independent determinations 
of atmospheric parameters over wide ranges of all atmospheric parameters is desirable to validate the synthetic spectra. 
For the empirical templates, on the other hand, both the spectra themselves and the atmospheric parameters associated 
 with the stars, the latter often determined with 
a variety of techniques, have uncertainties, which eventually propagate into the parameters estimated for target stars.   
To obtain robust atmospheric parameters for large samples of stars targeted by large spectroscopic surveys, 
it seems logical to use a subset of the survey spectra whose atmospheric parameters have been 
previously accurately determined by other means as the template or training set to obtain parameter estimates 
for the remaining spectra, considering that in this case, both the template (training) set of spectra and target spectra 
are obtained with the same instrument and thus likely have the same error patterns. However, to obtain 
accurate stellar parameters by independent means for a substantial subset of spectra covering 
wide ranges of parameters is an extremely time-consuming and 
challenging task yet to be accomplished for the individual completed/on-going surveys. 

For the LAMOST spectroscopic surveys, although a comprehensive subset of stellar spectra with accurately known 
atmospheric parameters is still absent, there are, however, thousands of stars targeted by LAMOST that have either 
accurate log\,$g$ measurements from asteroseismic analysis of the {\em Kepler} data \citep{Huber+2014}, 
or precise measurements of metallicity [M/H], $\alpha$-element to metal abundance ratio [$\alpha$/M] 
and individual elemental abundances [X/H] for elements C, N, O, Na, Mg, Al, Si, S, K, Ca, Ti, V, Mn, Fe and Ni 
determined from the APOGEE high resolution spectra \citep{Holtzman+2015}. There are also thousands of stars that have 
accurate Hipparcos parallax measurements \citep{Perryman+1997}, thus their accurate luminosities 
can be derived.  
Those several sets of stars are therefore useful as training sets for the estimation of either stellar 
atmospheric parameters, luminosities or  
individual elemental abundances from the LAMOST spectra.
Asteroseismic log\,$g$ estimates inferred from the {\em Kepler} data can be accurate to 0.03\,dex \citep{Hekker+2013, Huber+2014}, 
much better than achievable even with high-resolution spectroscopy ($\sim$0.1\,dex), though 
log\,$g$ values estimated from spectroscopy do not necessarily match fully with the asteroseismic results. 
The Hipparcos parallax measurements for stars in the solar neighborhood are accurate to a few per cent, 
corresponding to an accuracy of absolute magnitude better than about 0.2\,mag.   
The APOGEE spectra have a resolution about 22 500 \citep{Majewski+2010}, much higher than that of 
the LAMOST spectra ($\sim$1800). Estimates of [M/H], [$\alpha$/M] and individual elemental abundances 
yielded by APOGEE spectra should be accurate/precise enough to be used for the LAMOST stellar parameter 
estimation. 

Another important source of error for stellar atmospheric parameter determination comes 
from the inadequacy of algorithms used. Properties of an observed stellar spectrum 
are governed by the combination of a number of atmospheric parameters 
($T_{\rm eff}$, [Fe/H], [$\alpha$/Fe], log\,$g$, etc.), thus parameters estimated from the spectrum 
are often degenerated, especially for a low-resolution spectrum 
where spectral features are often blended. Compared to $T_{\rm eff}$, some parameters, 
in particular $\log\,g$, and some elemental abundances, are  
less sensitive to the observed spectral features. Estimates of those parameters therefore often suffer from lower 
accuracies if one uses the same metric (e.g. $\chi^2$) in pixel space 
of spectral flux for the estimation. The situation can be improved if specific 
spectral features can be singled out and used to estimate those parameters. 
However, singling out specific features sensitive to the individual parameters 
is not straightforward, due to the degeneracy of parameters as well as spectral 
blending, especially under low-resolution. The `specific features' , 
if exist, are likely to be non-linearly correlated with the observed flux in pixel space such 
that a non-linear algorithms is needed to find them out.  

In this paper, we explore a multivariate regression method based on 
kernel-based principal component analysis (KPCA) to estimate atmospheric parameters,  
absolute magnitudes, and individual elemental abundances from the LAMOST spectra, 
utilizing the aforementioned spectral training sets. 
KPCA is a non-linear method to extract features from high-dimension data sets. 
It was first proposed by \citet{Scholkopf+1998}, and validated by a series of 
work \citep[e.g.][]{Muller+2001, Zhang+2005}. Compared with the traditional (linear) PCA, 
KPCA can extract non-linear components, thus one expects that it may yield higher 
accuracy for the estimation of stellar atmospheric parameters. For the purpose, 
we have defined four sets of training spectra consisting of: the MILES spectral library 
template stars, the LAMOST-Hipparcos common stars, 
the LAMOST-$Kepler$ common stars and the LAMOST-APOGEE common stars. 
We present a detailed analysis of the precisions of parameters yielded by  
these training sets, and estimate the parameter 
uncertainties as a function of signal-to-noise ratio (SNR) and stellar atmospheric parameters. 
The newly developed method will be incorporated into the LAMOST Stellar Parameter 
Pipeline at Peking University \citep[LSP3;][]{Xiang+2015a}, and the resultant   
parameters for stars of the LAMOST Spectroscopic Survey of the Galactic Anticentre 
will be publicly available in the second release of value-added catalogues of LSS-GAC 
(LSS-GAC DR2; Xiang et al. 2016, in preparation).  
 
The paper is arranged as follows. In \S{2}, we briefly introduce the LAMOST Galactic surveys, 
data processing and parameter determination. 
In \S{3}, we introduce our new method to estimate stellar parameters. 
The training sets are defined in \S{4}. In \S{5}, we present results applying 
the method to spectra collected by the LSS-GAC, including a detailed error analysis. 
In \S{6}, we discuss briefly the potential future improvements of the method. 
This is followed by conclusions in \S{7}.

\section{The LAMOST Galactic Surveys}
The Large Sky Area Multi-Object Fiber Spectroscopic Telescope \citep[LAMOST, also named as the 
"Guo Shoujing Telescope";][]{Wang+1996, Cui+2012} is a $Wang-Su$ type reflecting Schmidt 
telescope that has both a wide field (5$\degr$ in diameter) and a large effective aperture (4 -- 6\,m, depending on the pointing altitude and hour angle). 
It collects simultaneously 4000 fiber spectra at a resolving power $R\sim1800$ (with slit masks of 2/3 
the fiber diameter of 3.3\,arcsec) of wavelength range 3700 -- 9000{\AA}.      

The LAMOST Galactic spectroscopic surveys -- LAMOST Experiment for Galactic Understanding 
and Exploration \citep[LEGUE;][]{Zhao+2012, Deng+2012, Liu+2015}, consist of three main components,  
the spheroid \citep{Deng+2012}, disk \citep{Hou+2013} and Anti-center \citep[LSS-GAC;][]{Liu+2014, Yuan+2015a} surveys. 
In addition to the main surveys, there are projects targeting specific sky areas, such as the LAMOST-$Kepler$ fields. 
The surveys aim to collect up to ten million stellar spectra down to 17.8 magnitude in the SDSS $r$-band 
(18.5 mag for limited fields). By June 2015, more than 5 million spectra have been obtained with SNRs higher than 10, 
the minimum required for a successful exposure \citep{Liu+2014, Yuan+2015a, Luo+2015}.  

The raw 2-dimension data are processed with the LAMOST 2D pipeline for spectral extraction, wavelength calibration, 
flat fielding, background subtraction and flux calibration to produce 1D spectra \citep{Luo+2015}.  
Given that the Anti-center survey (LSS-GAC) targets fields of low Galactic latitudes that suffer from 
high dust extinction, a specific flux calibration pipeline has been developed at Peking University 
for accurate spectral flux calibration \citep{Xiang+2015b}. 

Two pipelines have been developed to derive stellar parameters, including radial velocity 
$V_{\rm r}$, effective temperature $T_{\rm eff}$, surface gravity log\,$g$ and metallicity [Fe/H] from LAMOST 
spectra. The official LAMOST Stellar Parameter Pipeline \citep[LASP;][]{Wu+2011, Wu+2014, Luo+2015} 
estimates parameters with a $\chi^2$ minimization scheme developed based on the ULySS procedure \citep{Koleva+2009, Wu+2011, Wu+2014},  
utilizing the ELODIE spectral library \citep{Prugniel+2007} as the templates. 
The LAMOST Stellar Parameter Pipeline at Peking 
University \citep[LSP3;][]{Xiang+2015a, Ren+2016} determines atmospheric parameters 
by template matching with the MILES spectral library \citep{Sanchez-Blazquez+2006}, utilizing both 
a $\chi^2$-based weighted-mean and a $\chi^2$-minimization algorithm. 
Stellar atmospheric parameters of the MILES template stars have been (re-) homogenized by 
Huang et al. (2016, in preparation; cf. \S{4.1}). A major observational campaign is currently under way 
to expand the MILES library (Wang et al. 2016, in preparation), by observing additional template 
stars in order to improve the coverage and homogeneity of distribution of template stars in parameter space, 
as well as to extend the wavelength coverage of template spectra to the far red ($\sim 9000$\,{\AA}).    
By fixing $T_{\rm eff}$, log\,$g$ and [Fe/H] yielded by the $\chi^2$-based weighted-mean 
algorithm, LSP3 also determines [$\alpha$/Fe] by template 
matching with Kurucz synthetic spectral library \citep{Liji+2016}. Stellar parameters yielded by LASP are 
publicly available from the LAMOST official data release\footnote{http://dr1.lamost.org} \citep{Luo+2012, Luo+2015}. 
Currently, LSP3 is only applied to spectra collected by the LSS-GAC survey, and the resultant parameters, 
as well as additional parameters such as interstellar dust extinction and distance, estimated based on 
the LSP3 atmospheric parameters with various methods, are released as 
value-added catalogues\footnote{http://lamost973.pku.edu.cn/site/data} of LSS-GAC \citep{Yuan+2015a}.   
Apart from results from the above two stellar parameter pipelines, \citet{Liuchao+2015} use
a support vector regression (SVR) method to estimate log\,$g$ from LAMOST spectra for giant stars, 
taking the LAMOST-$Kepler$ stars with asteroseismic log\,$g$ measurements as the training set. 
Ho et al. (2016) estimate stellar atmospheric parameters ($T_{\rm eff}$, log\,$g$, 
[Fe/H] and [$\alpha$/Fe]) for giant stars in the LAMOST DR2 by tying the LAMOST spectra to the 
APOGEE DR12 estimates of stellar parameters with {\em the} $Cannon$ \citep{Ness+2015}.  

\section{the method}
\subsection{The Kernel-based Principal Components Analysis}
A detailed introduction of the Kernel-based PCA can be found in \citet{Scholkopf+1998} and \citet{Muller+2001}. 
Below we briefly summarize the algorithm for completeness. 
Let $\bm{x}_k$, $k = 1, ..., M$, denote the spectra of $m$ stars, each contains $N$ wavelength pixels, 
$\bm{x}_k \in {\rm R}^N$. The spectra are normalized such that the sum of all pixel squared values of a given spectrum is unity. 
Note that the normalization is critical to generate realistic values of the kernel function.
To extract data structures with KPCA, we map the spectra into feature space $F$ by a (nonlinear) function $\Phi(x_k)$.   
Then we have the covariance matrix in $F$, 
\begin{equation}
    C = \frac{1}{M}\sum_{i=1}^M\Phi(\bm{x}_i)\Phi(\bm{x}_i)^{\rm T}. 
\end{equation}
To calculate the principal components, we solve the Eigenvalue problem below to find Eigenvalue 
$\lambda > 0$ and Eigenvector $\bm{V}\neq0$: 
\begin{equation}
    \lambda\bm{V} = C\bm{V} =\frac{1}{M}\sum_{i=1}^M(\Phi(\bm{x}_i)\cdot\bm{V})\Phi(\bm{x}_i).
\end{equation}
All Eigenvectors with nonzero Eigenvalue can be written in the span of $\Phi(\bm{x}_1)$, ..., $\Phi(\bm{x}_M)$ 
such that,  
\begin{equation}
\bm{V} = \sum_{j=1}^{M}\alpha_j\Phi(\bm{x}_j).
\end{equation} 
Multiplying Eq. (2) by $\Phi(\bm{x}_k)$ from the left yields,  
\begin{equation}
\begin{aligned}
&\lambda\sum_{j=1}^{M}\alpha_j(\Phi(\bm{x}_k)\cdot\Phi(\bm{x}_j)) = \\
&\frac{1}{M}\sum_{j=1}^M\alpha_j\sum_{i=1}^{M}(\Phi(\bm{x}_k)\cdot\Phi(\bm{x}_i))(\Phi(\bm{x}_i)\cdot\Phi(\bm{x}_j)).
\end{aligned}
\end{equation}    
Defining an $M \times M$ matrix $K$,  
\begin{equation}
K_{ij} := (\Phi(\bm{x}_i)\cdot\Phi(\bm{x}_j)),
\end{equation} 
the Eigenvalue problem becomes,  
\begin{equation}
M\lambda\bm{\alpha} = K\bm{\alpha}, 
\end{equation}   
where $\bm{\alpha} = (\alpha_1, ..., \alpha_M)$. 
The solutions of the Eq. (6) are normalized by imposing $\lambda_k(\bm{\alpha}^k\cdot\bm{\alpha}^k) = 1$. 
The normalization yields $(\bm{V}^k\cdot\bm{V}^k) = 1$ for all $k=p, ..., M$, where $p$ is the index 
corresponding to the first nonzero Eigenvalue $\lambda_p$.    
The data in $F$ are centered by substituting $K$ with,  
\begin{equation}
\hat{K}=K-1_MK-K1_M+1_MK1_M, 
\end{equation}
where $(1_M)_{ij} = 1/M$.

To extract the principal components, we calculate the projections of the spectra on Eigenvectors $\bm{V}^k$ in $F$. 
For a given test spectrum $\bm{x_t}$, 
the corresponding nonlinear principal components are   
\begin{equation}
 (\bm{V}^k\cdot\Phi(\bm{x_t})) = \sum_{j=1}^M\alpha_j^k(\Phi(\bm{x}_j)\cdot\Phi(\bm{x_t})).
\end{equation}

Even in the most realistic cases, the nonlinear transformation $\Phi$ in general can not be expressed explicitly. 
Therefore, instead of calculating the products $(\Phi(\bm{x})\cdot\Phi(\bm{y}))$ in Eq. (6) directly, we 
use a kernel representation of the form, 
\begin{equation}
k(\bm{x},\bm{y}) =  (\Phi(\bm{x})\cdot\Phi(\bm{y})). 
\end{equation}
Various forms of kernel function, such as polynomial, radial basis functions and 
sigmoidal, as well as other more complicated kernels, have been validated \citep{Scholkopf+1998, Muller+2001}. 
In the current work, we use the Gaussian radial basis functions,  
 \begin{equation}
 k(\bm{x},\bm{y}) = {\rm exp}(\frac{-\|\bm{x}-\bm{y}\|^2}{c}), 
 \end{equation}
where $\|\cdot\|$ represents the Euclidean norm, $\|\bm{x}-\bm{y}\| \equiv \sqrt{(\bm{x}-\bm{y})\cdot(\bm{x}-\bm{y})}$, 
and $c$ is the width of the kernel. Throughout this paper, we adopt $c = 0.005$, a typical value of 
the squared Euclidean norm $\|\bm{x}-\bm{y}\|^2$ for the LAMOST spectra. 
In fact, we have examined different values of $c$ (e.g. 0.005, 0.05, 0.5, 1.0, 5.0) making 
use of both LAMOST-$Kepler$ sample stars and member stars of open clusters, and 
found that 0.005 is an optimal one.

\subsection{The Regression}
To derive atmospheric parameters from the principal components, we construct a multiple-linear 
relation between the principal components $\bm{P}$ and the stellar atmospheric parameters 
for each parameter $\bm{y}$, 
\begin{equation}
\bm{y} = \sum_{i=1}^Nc_i\bm{P}_i + c_0,
\end{equation} 
where $N$ is the adopted number of principal components, determined empirically with 
a brute-force search (cf. \S{4}), $c_0$ is a constant, 
and $\bm{y}$ is any one of the parameters $T_{\rm eff}$, log\,$g$, [Fe/H], ${\rm M}_V$, ${\rm M}_{K_{\rm s}}$, 
[M/H], [$\alpha$/M], [$\alpha$/Fe], [C/H] and [N/H].
The coefficients $c_i, i=1,...,N,$ are determined by a least square multiple-linear fit to a training data set. 

When estimating log\,$g$ values for giant stars using the LAMOST-$Kepler$ sample stars as the training set (cf. \S{4.3}), 
$T_{\rm eff}$ and [Fe/H] from LSP3 are adopted as priors. This is carried out by taking log\,$T_{\rm eff}$ 
and [Fe/H] as input pixel values of spectral flux. Similarly, when estimating log\,$g$ values for dwarfs 
using the MILES library as the training set (cf. \S{4.1}), as well as when estimating [M/H], [$\alpha$/M] and [$\alpha$/Fe]
using the LAMOST-APOGEE stars as the training set (cf. \S{4.4}), $T_{\rm eff}$ yielded by LSP3 is adopted 
as a prior. For the estimation of individual elemental abundances [Fe/H], [C/H] and [N/H] with the 
LAMOST-APOGEE training set, the LSP3 $T_{\rm eff}$ as well as the KPCA estimated [M/H] are 
adopted as the priors. Note however that, the priors are found to have only minor effects on the parameter 
estimation. The method remains robust even if incorrect values of $T_{\rm eff}$ and [Fe/H] are provided. 
This is probably due to the fact that LAMOST spectra contain sufficient information (features) 
to yield correct log\,$g$ values and elemental abundances without relying on those priors. 
This also implies that our method is robust enough and is largely free from degeneracy of various parameters. 
Since the priors are not very helpful to improve the parameter estimation, for the estimation of 
absolute magnitudes using the LAMOST-Hipparcos stars as the training set (cf. \S{4.2}) as well as 
the estimation of $T_{\rm eff}$ and [Fe/H] using the MILES stars as the training set (cf. \S{4.1}), 
no priors are used for simplicity.

\subsection{Preprocessing for the spectra}
Though the LAMOST spectra cover a wavelength range of 3700 -- 9000\,{\AA}, 
we have opted to use the 3900 -- 5500\,{\AA} segment only for the stellar parameter determination  
with the following considerations. Firstly, for the majority of stars in our sample, except for those 
very red ones, spectral features are mostly found in the blue part of the spectra, 
while the red part of the spectra contains much less features. The red part spectra 
also suffer from serious background contamination, including sky emission lines and 
telluric bands. Secondly, due to the low instrument efficiency near the 
edges of wavelength coverage of the blue- and red-arm spectra, for a considerable fraction of stars, 
the blue- and red-arm spectra are not perfectly jointed together but show artificial discontinuities. 
This is particularly serious for very bright stars observed under grey and bright lunar conditions 
and for faint stars of low spectral SNRs. Finally, the wavelength coverage of 
MILES spectra reach only 7410\,{\AA} in the red. Note that the very blue part (3700 -- 3900\,{\AA}) 
of the spectra is discarded because of the low spectral SNRs in this spectral regime for the majority of stars. 

Given the relatively large uncertainties of spectral flux calibration of LAMOST spectra 
and the unknown extinction values for most of the survey targets, especially those  
within the Galactic disc, it is essential to use continuum-normalized spectra to deliever reliable 
stellar parameters. A reasonable continuum normalization is found to be crucial for robust estimation 
of stellar parameters. Since the MILES spectra are generally 
better calibrated and have higher SNRs than most LAMOST spectra, pesudo-continua determined for the former 
are in general more accurate. We therefore adopt the pesudo-continua determined for MILES spectra as a 
starting point to derive pesudo-continua for LAMOST spectra. Specifically, the pesudo-continua of MILES 
spectra are first derived using a 5th-order polynomial for the wavelength range 3850 -- 5800\,{\AA}.  
The pesudo-continuum of a LAMOST spectrum is then derived by scaling that of the 
best-matching MILES spectrum as yielded by LSP3 with a 5th-order polynomial. 
Here the 5th-order polynomial is introduced to account for any differences of pesudo-continua
between the LAMOST spectrum and the best-matching MILES spectrum. 
Such differences can be intrinsic or arise from differences in the interstellar extinction or are simply   
artifacts induced by inadequacies in the data reduction (e.g. uncertainties of flux calibration).       

It is found that some LAMOST spectra have bad pixels of unrealistic 
fluxes as a consequence of inadequacy in the data reduction (e.g. poor cosmic ray removal). 
The presence of those bad pixels could have dramatic adverse effects on parameter estimation (cf. \S{5.1}). 
To avoid the problem, we calculate the differences between the normalized LAMOST and best-matching MILES spectra, 
and identify bad pixels with 10$\sigma$ deviations. The spectral fluxes 
of bad pixels are replaced by values of Bessel interpolation of nearby pixels. 

\section{The training data sets}
We define four training data sets in this work: the MILES spectral library, the LAMOST-Hipparcos 
common stars, the LAMOST-$Kepler$ 
stars with asteroseismic measurements of log\,$g$ and the LAMOST-APOGEE common stars. 
The LAMOST-$Kepler$ stars are used to estimate log\,$g$ only, and the LAMOST-APOGEE stars 
are used to estimate metallicity [M/H], $\alpha$-element to metal abundance ratio [$\alpha$/M], 
$\alpha$-element to iron abundance ratio [$\alpha$/Fe] and elemental abundance [Fe/H], [C/H] and [N/H].  
Both the LAMOST-$Kepler$ and LAMOST-APOGEE data 
sets are used to estimate parameters for giant stars only given that they contain few dwarfs.  
The MILES spectral library is used to estimate $T_{\rm eff}$ and [Fe/H] for all stars, and 
log\,$g$ for dwarf and subgiant stars. Note that [Fe/H] deduced using the 
MILES spectra as the training set are also referred to metallicity throughout the paper. 
The LAMOST-Hipparcos common stars, combined with the MILES 
stars with accurate measurements of parallax, are used to estimate absolute magnitudes. 
A summary of the adopted training sets and their effective parameter ranges, as 
well as the number of principle components adopted for the regression, are 
presented in Table\,1. Note that for giant stars, there are two estimates 
of [Fe/H], which are estimated based on the MILES training set and the 
LAMOST-APOGEE training set, respectively.
\begin{table*}
\centering
\caption{Adopted training sets, number of PCs and effective parameter ranges.}
\label{}
\begin{tabular}{llll}
\hline
 Training set              & $N_{PC}$                    & Parameters            & Effective parameter range            \\
\hline
MILES    &  100                               &  $T_{\rm eff}$, [Fe/H]           &   $3000 < T_{\rm eff}^{\rm LSP3} < 12000$\,K \\
               &                                       &   log\,$g$                              &   $3000 < T_{\rm eff}^{\rm LSP3} < 12000$\,K \& log\,$g_{\rm LSP3}$ > 3.0 \\
LAMOST-Hipparcos stars & 100     &    ${\rm M}_V$, ${\rm M}_{K_{\rm s}}$  &  $3000 < T_{\rm eff}^{\rm LSP3} < 12000$\,K \\
LAMOST-$Kepler$ stars  &  90       &  log\,$g$                               & $T_{\rm eff}^{\rm LSP3} < 5500$\,K  \& log\,$g_{\rm LSP3}$ < 3.8  \\   
LAMOST-APOGEE stars &   90      & [M/H], [Fe/H], [$\alpha$/M], [$\alpha$/Fe], [C/H], [N/H] & $T_{\rm eff}^{\rm LSP3} < 5500$\,K  \& log\,$g_{\rm LSP3}$ < 3.8  \\    
\hline
\end{tabular}
\end{table*}

\subsection{The MILES spectral library}
\begin{figure*}
\centering
\includegraphics[width=160mm]{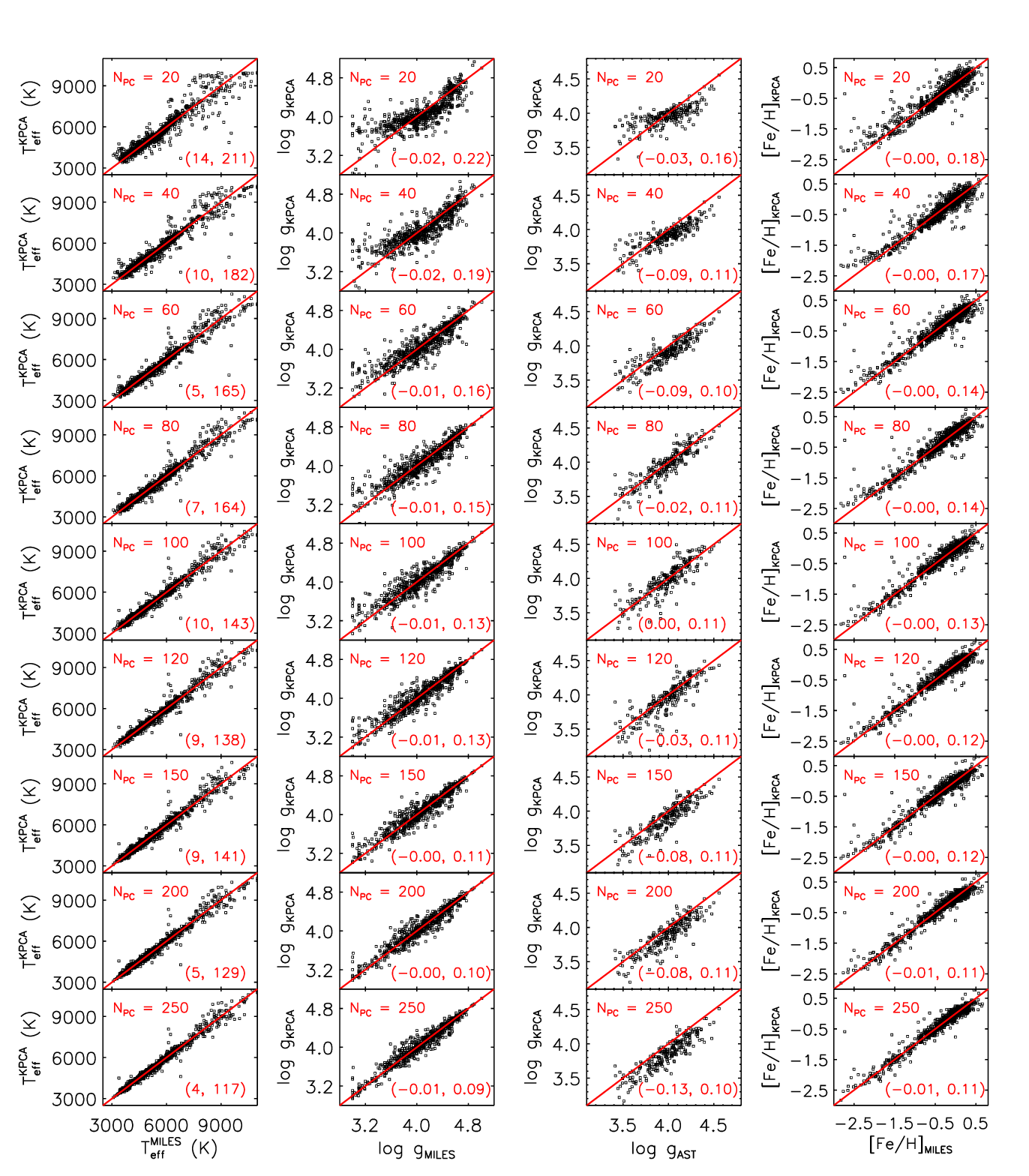}
\caption{Comparison of KPCA atmospheric parameters with the MILES values (Columns 1, 2 and 4 
 for $T_{\rm eff}$, log\,$g$ and [Fe/H], respectively.) for the MILES training stars. 
 The KPCA $\log\,g$ estimates are compared to the asteroseismic measurements log\,$g_{\rm AST}$ 
 in Column 3 for the LAMOST-$Kepler$ stars of log\,$g_{\rm AST}>3.4$\,dex. For each column, different 
panels show results for different number of principal components adopted. 
The number of principal components, as well as a resistant estimates of the mean and 
standard deviation of the parameter differences, are marked in the plots.}
\label{Fig1}
\end{figure*}
The MILES spectral library includes 985 stars covering parameter space $3000<T_{\rm eff}<40000$\,K, 
$0<{\rm log}\,g<5$\,dex and $-3.0<{\rm [Fe/H]}<0.5$\,dex. The wavelength coverage of the spectra
is 3525 -- 7410\,{\AA}, and the spectral resolution is about 2.5\,{\AA} \citep{Sanchez-Blazquez+2006, Falcon-Barroso+2011}. 
The latter is close to that of the LAMOST spectra.  
The MILES spectra were taken with a long-slit spectrograph, and specific efforts were carried out for 
accurate flux calibration \citep{Sanchez-Blazquez+2006}, yielding stellar spectral energy distributions (SEDs)  
more reliable than possible with high-resolution echelle spectroscopy. 
The parameters of the MILES stars are collected from the literatures, mostly determined with high-resolution 
spectroscopy. \citet{Cenarro+2007} have carried out a homogenization of the collected 
parameters to correct for systematics amongst values from different sources. 
Nevertheless, the homogenization of \citet{Cenarro+2007} was carried out for a 
limited range of temperature, $4000<T_{\rm eff}<6300$\,K, and parameter values used for  
the template stars were obtained in early time (e.g. before 1990), with relatively large random errors. 
To reduce both the systematic and random errors of parameters of the MILES stars, 
Huang et al. (2016, in preparation) have recompiled usable determinations of parameters 
of the stars by replacing old measurements with more recent ones available from the PASTEL catalog \citep{Soubiran+2010}. 
The compiled values of effective temperature are then determined using the metallicity-dependent colour-temperature 
relations deduced from more than two hundred nearby stars with direct effective 
temperature measurements \citep{Huang+2015b}. Note that Huang et al.'s relations 
are only available for stars of $T_{\rm eff}<10$\,000\,K and ${\rm [Fe/H]}>-1.0$\,dex. For stars outside 
those parameter ranges, the compiled temperatures are adopted. Values of surface gravity are 
re-determined using Hipparcos parallax \citep{Perryman+1997, Anderson+2012} 
and stellar isochrones from the Dartmouth Stellar Evolution Database \citep{Dotter+2008}.   
Finally, values of metallicity are re-calibrated to the scale of Gaia-ESO survey \citep{Jofre+2015}.
Here, we adopt those re-calibrated/determined  parameters. Note that for the majority of stars, 
the re-calibrated/determined parameters are consistent well with the values of \citet{Cenarro+2007}.

MILES stars of $T_{\rm eff} \le 12000$\,K are selected as our training set 
for the estimation of $T_{\rm eff}$ and [Fe/H] for LAMOST stars.  
Stars of $T_{\rm eff} > 12000$\,K were excluded from the training set because 
tests show that if they were included, residuals of the multi-linear regression for 
F/G/K stars became very large.  
Here we concentrate on A/F/G/K stars, and leave the parameter determination 
for stars of earlier-types a future work. 
Tests also show that excluding giant stars (log\,$g < 3.0$\,dex) 
in the training set yields significantly better results for dwarf and subgiant stars than otherwise. 
We therefore use MILES stars with log\,$g > 3.0$\,dex only in the training sample 
since we want to determine accurate log\,$g$ for dwarf and subgiant stars.

\begin{figure*}
\centering
\includegraphics[width=160mm]{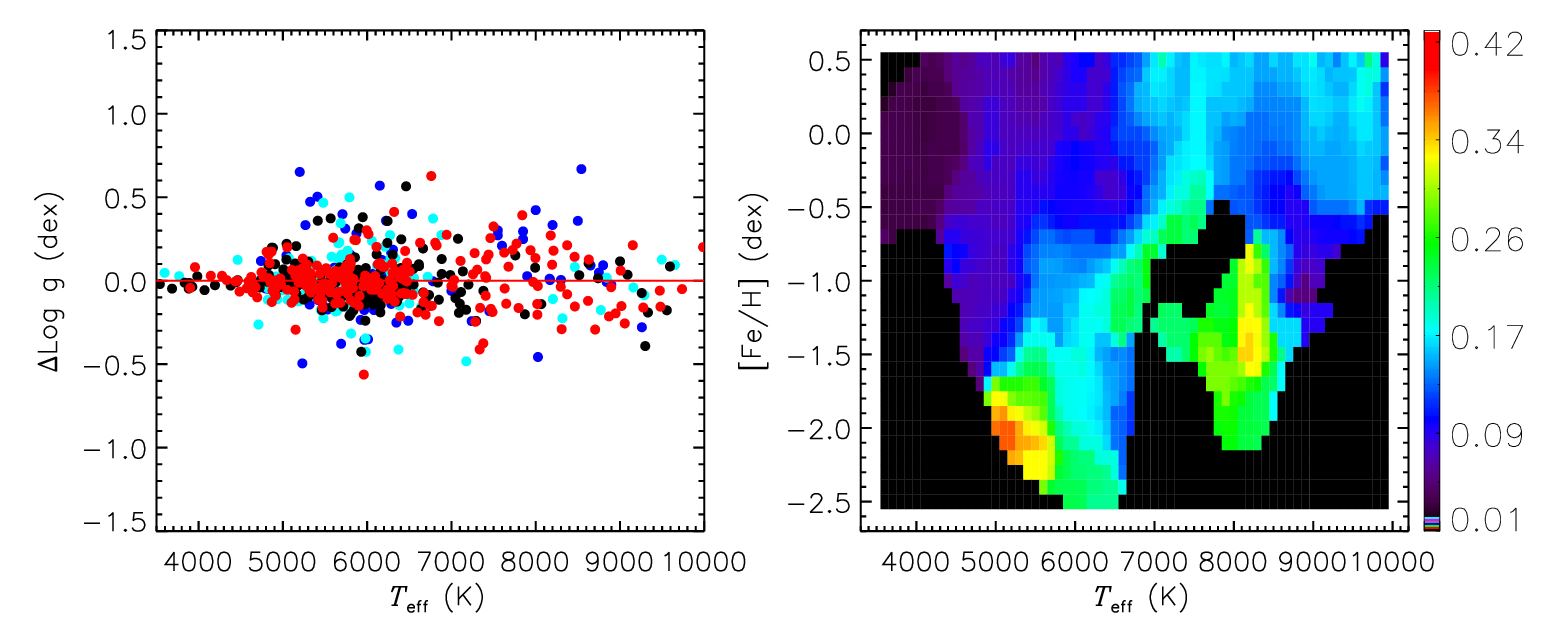}
\caption{Residuals of regression of log\,$g$ estimation utilizing the MILES library 
as the training data set. $Left$: Residuals of log\,$g$ of MILES stars in different 
metallicity bins (red: ${\rm [Fe/H]}>0$; black: $-0.5<{\rm [Fe/H]}<0$; cyan: $-1.0<{\rm [Fe/H]}<-0.5$; 
blue: ${\rm [Fe/H]}<-1.0$\,dex); 
$Right$: Distribution of the dispersions of log\,$g$ residuals in the $T_{\rm eff}$ -- [Fe/H] plane.}
\label{Fig2}
\end{figure*}

To select an optimal value for the number of principal components $N_{\rm PC}$ used for the regression, we  
compare the original MILES parameters with values deduced from the KPCA using different values of 
$N_{\rm PCs}$. The results are plotted in Fig.\,1.   
The Figure also compares the deduced log\,$g$ of the LAMOST-$Kepler$ 
dwarf/subgiant stars with the asteroseismic measurements. 
The Figure shows that as $N_{\rm PC}$ increases, the agreement generally improves. 
As $N_{\rm pc}$ increases from 20 to 100, the dispersion of residuals of $T_{\rm eff}$ 
decreases from 211\,K to 143\,K, that of log\,$g$ decreases from 0.22\,dex to 0.13\,dex, 
and that of [Fe/H] decreases from 0.18 to 0.13\,dex. 
Note that here the dispersion is a resistant estimate of the standard deviation.
The comparisons of KPCA log\,$g$ values deduced from the LAMOST 
spectra of the LAMOST-$Kepler$ stars with asteroseismic log\,$g$ measurements show that the systematic 
difference reaches a minimum of 0.00\,dex at $N_{\rm PC} = 100$, and the dispersion is also small (0.11\,dex). 
Although not plotted here, we have also carried out similar tests using candidate member stars
of open clusters M\,67, that have a significant number of member candidates targeted by the LAMOST. 
It is found that at an $N_{\rm PC}$ value around 100, the resultant $T_{\rm eff}$ -- log\,$g$ diagrams of  
cluster member star candidates match well with the theoretical isochrones. 
Based on the above tests, we have adopted $N_{\rm PC}=100$ for the MILES training set. 

Note that though Fig.\,1 shows that $N_{\rm PC}=100$ is the best for log\,$g$ estimation, 
it is not necessary the best choice for the estimation of $T_{\rm eff}$ and [Fe/H] because the
training samples are different --- all MILES stars are used to estimate the latters, 
while only those of log\,$g > 3.0$ are used 
to estimate log\,$g$. Nevertheless, we have adopted $N_{\rm PC}=100$ for the estimation 
of $T_{\rm eff}$ and [Fe/H] considering that for this $N_{\rm PC}$ value, the regression residuals 
are already quite small and acceptable. The selection of an optimal number of $N_{\rm PC}$ 
if an effort to find a balance: although a larger value of $N_{\rm PC}$ will generate  
smaller residuals of the regression relations, an excessively large $N_{\rm PC}$  
will also induce some undesired features caused by imperfections of the spectra. 
The latter could be potentially serious given that the training spectra (MILES) and the target 
spectra (LAMOST) are obtained with different instruments. 
In addition, it is found that the optimal value of $N_{\rm PC}$ varies with the 
adopted value of kernel width $c$. In this paper, we have adopted a fixed value of $c$ 
(0.005). However, it is found that if we adopt a $c$ value of 0.5, the optimal 
value of $N_{\rm PC}$ becomes $\sim$\,30.

Fig.\,2 plots the variations of regression residuals of log\,$g$ estimates as functions of 
$T_{\rm eff}$ and [Fe/H]. 
The left panel shows the residuals of log\,$g$ of stars in different [Fe/H] bins as a function 
of $T_{\rm eff}$. It shows that the residuals of hot stars ($T_{\rm eff} > 7000$\,K) as well as 
metal-poor stars have larger dispersions than cooler, metal-rich stars. 
The dispersion is a measure of the precision of log\,$g$ measurements of the MILES stars.
The right panel shows the distribution of dispersions in the $T_{\rm eff}$ -- [Fe/H] plane. 
To calculate the dispersions, we first create a dense grid of $T_{\rm eff}$ and [Fe/H], with a 
step of 100\,K and 0.1\,dex in $T_{\rm eff}$ and [Fe/H], respectively. 
At each grid point, we search for the nearby training stars located within a box of size  
100\,K by 0.1\,dex. If the number of stars within the box is smaller 
than 20, we increase the box size by 50\,K in $T_{\rm eff}$ and 0.05\,dex 
in [Fe/H] until the box size reaches an upper limit of 500\,K by 0.5\,dex. Finally, if the number of stars 
in the box is larger than 5, then the mean and standard deviation are calculated in a 
resistant way. Note that values of the dispersions shown in the Figure are 
square root of the mean of squares, i.e. the square of the mean and the square of the standard deviation. 
Here the square root of the mean of squares is preferred rather than the standard deviation 
because for some places of the parameter space 
(e.g. metal-poor side) or for bins where the numbers of stars are small, even though 
the computed standard deviation is small, the mean residual can be large. 
The Figure shows that the dispersions vary significantly 
over the $T_{\rm eff}$ -- [Fe/H] plane. For $T_{\rm eff}<7000$\,K 
and ${\rm [Fe/H]} > -1.0$\,dex, the dispersions can be as small as $\sim$0.1\,dex, indicating 
a high precision of log\,$g$ determination for the F/G/K-type metal-rich stars. 
For metal-poor (${\rm [Fe/H]} < -1.0$\,dex) or relatively hot ($T_{\rm eff}>7000$\,K) 
stars, the dispersions are significantly larger. Such an error pattern is consistent with the results 
shown in Fig.\,1, where the dispersions of the overall residuals of MILES stars could be 
significantly larger than those deduced for the LAMOST-$Kepler$ stars, since the latter  
consist entirely of F/G stars with $5000 <T_{\rm eff} < 7000$\,K.    

\subsection{The LAMOST-Hipparcos stars}
The LAMOST has collected spectra of many very bright stars utilizing moon nights. 
A cross-identification of the LAMOST stars observed before Oct. 2015 with the Hipparcos 
catalog \citep{Perryman+1997} yields more than 7000 common stars. 
A considerable fraction of these Hipparcos stars have accurate measurements of parallax 
and magnitudes, such that their luminosity (absolute magnitude) can be derived with good 
accuracy. We take those stars as a training set to estimate directly stellar absolute magnitudes  
from the LAMOST spectra. To construct this data set, we require that the training 
stars have uncertainties of the absolute magnitudes, which are propogated from the 
parallax errors, smaller than 0.3\,mag, and we also require the training star have a LAMOST 
SNR higher than 50. These criteria lead to 810 stars left in our training data set. 

\begin{figure}
\centering
\includegraphics[width=80mm]{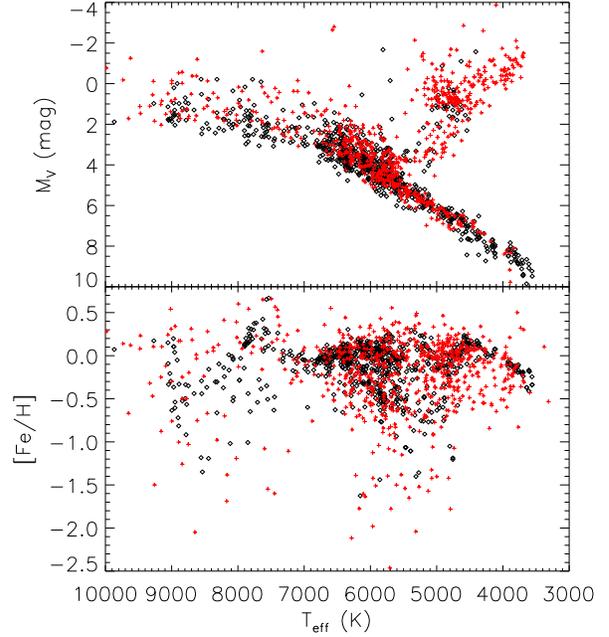}
\caption{Distribution of the luminosity training stars in the $T_{\rm eff}$ -- ${\rm M}_V$
and $T_{\rm eff}$ -- [Fe/H] plane. The black diamonds are the LAMOST-Hipparcos 
stars, while the red plus are the MILES stars. $T_{\rm eff}$ and [Fe/H] of 
the LAMOST-Hipparcos stars are derived with the LSP3.}
\label{Fig3}
\end{figure}
Though most of the training stars are local enough so that the inter-stellar extinction are negligible, 
a few of them are found to suffer considerable extinction ($E_{B-V}\sim0.1$\,mag), 
and need to be corrected for. To derive extinction for those training stars, we deduce 
the intrinsic colour $(V-K_{\rm s})_0$ utilizing stellar atmospheric parameters derived from 
the LAMOST spectra with the LSP3 and the metallicity-dependent colour-temperature relation 
of \citet{Huang+2015b}. The distribution of the deduced $E_{B-V}$ has an overall peak at 0.01\,mag, 
with a dispersion of 0.03\,mag, as well as a small tail which corresponds to stars with significant extinction.     
We therefore correct for extinction for 161 stars both of $E_{B-V} > 0.03$\,mag and 
with distance larger than 50\,pc. 

Considering the number of stars in the training set is small, and more important, there are 
rare metal-poor (${\rm [Fe/H]} < -1.0$) stars, we choose to add 742 MILES stars in 
the training set. Those MILES stars have also accurate parallax from the Hipparcos 
catalog, which result in an uncertainty in absolute magnitude smaller than 0.3\,mag.  
As a result, there are 1552 training stars in total for the estimation of stellar luminosity. 
The distribution of those training stars in the $T_{\rm eff}$ -- ${\rm M}_V$, 
$T_{\rm eff}$ -- [Fe/H] plane are shown in Fig.\,3, with the LAMOST-Hipparcos stars 
and the MILES stars shown in different symbols. 
 
\begin{figure*}
\centering
\includegraphics[width=140mm]{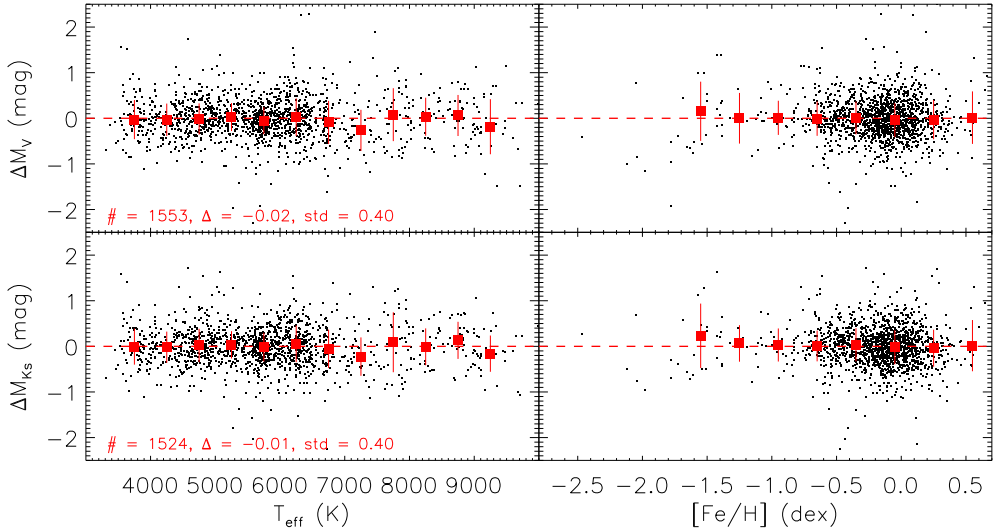}
\caption{Residuals of absolute magnitudes for the LAMOST-Hipparcos 
training stars as a function of effective temperature and metallicity. 
The upper panels are results for the Johnson $V$-band magnitude, while the 
lower panels are results for the 2MASS $K_{\rm s}$-band magnitude. 
The number of the training stars, as well as the mean and dispersion
of the distribution of the residuals estimated with a Gaussian fit, 
are marked in the plot.}
\label{Fig4}
\end{figure*}  

To select the optimal number of PC used for the regression, we have examined  
results deduced assuming different numbers of PC from 20 to 500 for a test data set, which contains 
both LAMOST-Hipparcos stars of SNR smaller than 50 and counterpart stars of duplicate observations 
of the training stars , as well as open cluster member star candidates. 
Finally, we adopt 100 PCs for the multiple linear regression. Figure\,4 shows the residuals 
of the regression for the Johnson $V$ and the 2MASS $K_{\rm s}$-bands absolute magnitudes 
as functions of effective temperature and metallicity. Here for the LAMOST-Hipparcos training 
stars, the temperature and metallicity are values yielded by LSP3. 
The Figure shows that though the dispersion of the residuals for stars of 
higher temperature or lower metallicity are slightly larger, there is no visible bias 
of the mean residuals respect to the temperature and metallicity. 
Values of the mean and standard deviation of the overall residuals are 
respectively $-0.02$ and 0.40\,mag for the $V$-band absolute magnitude, and
$-0.01$ and 0.40\,mag for the $K_{\rm s}$-band absolute magnitude. 

\subsection{The LAMOST-$Kepler$ stars}
The LAMOST-$Kepler$ project targets stars in the $Kepler$ fields with the LAMOST, 
and more than 100 000 spectra have been collected by September, 2014 \citep{De_Cat+2015}.
We have processed the raw LAMOST-$Kepler$ data collected since the beginning of the 
LAMOST pilot survey initiated in October, 2011 until June 2014,  using the LAMOST 2D pipeline 
\citep{Luo+2015} and the flux calibration pipeline developed for the LSS-GAC 
survey \citep{Xiang+2015b}. The data set includes about 53\,000 spectra of SNR higher than 10 
per pixel at 4650\,{\AA}. The stellar atmospheric parameters are derived from the spectra with LSP3.  
Amongst them, 3954 unique stars are found to have asteroseismic measurements of log\,$g$ 
from \citet{Huber+2014}. Most of them (3709) are giant stars of log\,$g_{\rm AST} < 3.4$\,dex. 
Only 245 stars have log\,$g_{\rm AST}$ values larger than 3.4\,dex. 

The common stars are divided into two samples, a training sample and a test sample. 
The training sample is used to generate regression relations between the principal components 
and the stellar parameters, while the test sample is used to evaluate the generated relations. 
To select stars of the training sample, we first discard stars with a spectral SNR lower than 50, then divide 
the remaining stars into small cells in the $T_{\rm eff}$ -- log\,$g$ -- [Fe/H] space. Here $T_{\rm eff}$ 
and [Fe/H] refer to values derived from the LAMOST spectra with LSP3, and log\,$g$ refers to  
the asteroseimic values. The training stars are then randomly selected from the individual cells. 
The selection is to ensure that the training stars are distributed as 
widely and homogeneously as possible in the parameter space. Half of the stars with 
a spectral SNR higher than 50 are selected as the training stars. The other half, together  
with those with a spectral SNR lower than 50, are adopted as the test sample. 
Although we intend to use this training set to estimate log\,$g$ for giant 
stars only (cf. \S{1}), it is found that keeping a sufficient number of dwarf stars in the training 
sample is necessary to avoid systematic errors of $\log\,g$ deduced for stars of log\,$g > 3.0$\,dex. \citet{Huang+2015a} 
shows that there is an artificial systematic trend in the KPCA estimates of log\,$g$ for stars of a log\,$g$ 
value larger than 3.0\,dex. The artifacts disappear if dwarf stars are also 
included in the training set. Considering that the number of dwarf stars are very small 
in our sample, we include all stars of log\,$g > 3.4$\,dex in our training set. 
In total, the training set contains 1520 stars. Fig.\,5 shows the distribution of the 
training sample in the $T_{\rm eff}$ -- log\,$g$ and log\,$g$ -- [Fe/H] planes. The Figure shows  
that most of the stars have a metallicity ${\rm [Fe/H]} > -1.0$\,dex, suggesting  
a disk sample.  

\begin{figure}
\centering
\includegraphics[width=80mm]{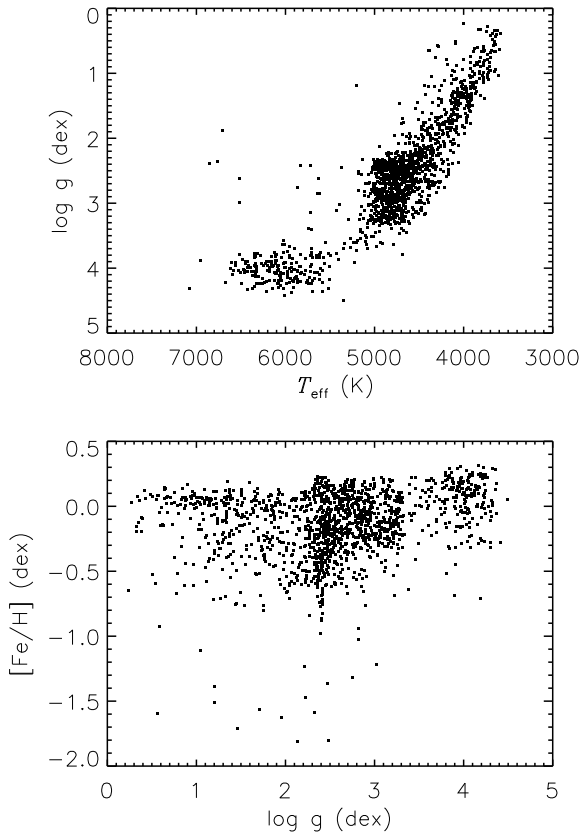}
\caption{Distribution of the LAMOST-$Kepler$ training stars in the $T_{\rm eff}$ -- log\,$g$ 
and log\,$g$ -- [Fe/H] planes. Values of $T_{\rm eff}$ and [Fe/H] are those yielded by LSP3, while 
those of log\,$g$ are asteroseismic measurements from \citet{Huber+2014}.}
\label{Fig5}
\end{figure}

\begin{figure*}
\centering
\includegraphics[width=160mm]{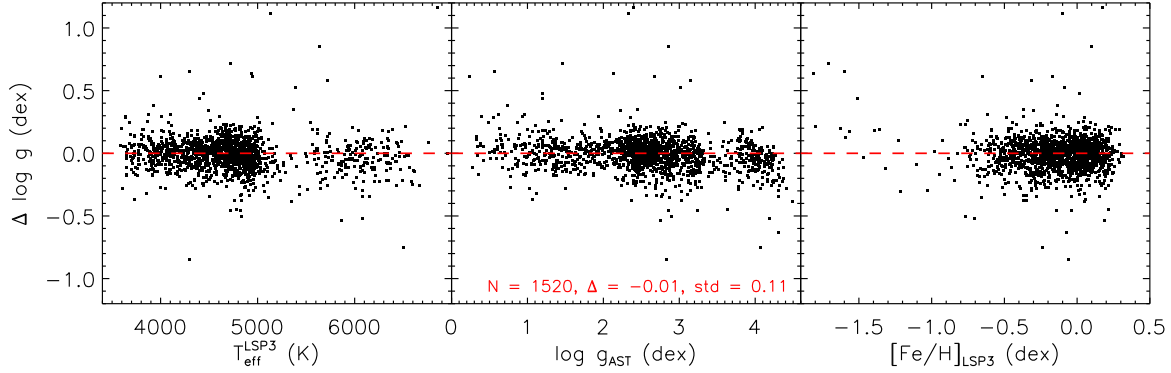}
\caption{Residuals of regression of log\,$g$ estimation utilizing the LAMOST-$Kepler$ 
stars as the training data set. The residuals are shown as a function of stellar parameters. 
The number of training stars, as well as resistant estimates of the mean and standard deviation 
of the residuals are marked in the middle panel.}
\label{Fig6}
\end{figure*}
Since this LAMOST-{\em Kepler} training set is only used for $\log\,g$ estimation, 
the values of $T_{\rm eff}$ and [Fe/H] of stars in the sample given by 
the LSP3 are adopted as prior to increase the robustness of results. 
To choose an optimal number of PCs for the multi-linear regression, we carry out 
tests similar to those done for the MILES training set. We have also applied different numbers of 
PCs to the test sample, and examine the robustness of the residuals. Finally, we adopt a 
number of 90 PCs and a kernel-width of 0.005. Note that the optimal number of PCs changes 
with the adopted kernel-width. Generally, it is found that for a larger kernel-width, a smaller 
number of PCs is required. For instance, if one adopts a the kernel-width of 0.5, a number 
of about 30 PCs seems to yield the optimal results. Fig.\,6 shows the regression residuals of log\,$g$  
of the training stars for a kernel-width $c=0.005$ and a number of PCs $N_{pc}=90$. The mean 
of the overall residuals is $-0.01$\,dex, with a dispersion of 0.11\,dex. The residuals show no significant 
trends with $T_{\rm eff}$, log\,$g$ and [Fe/H] for stars of log\,$g < 4.0$\,dex and ${\rm [Fe/H]} > -1.0$\,dex. 
Given that there are few of very metal-poor training stars, results for more metal-poor stars are probably  
less accurate. Values of $\log\,g$ for stars of log\,$g > 4.0$\,dex may have been underestimated. 
Since we focus on giant stars with this training set, we have ignored this potential systematics.  
   
\subsection{The LAMOST -- APOGEE common giant stars}

The SDSS-III/APOGEE collects $H-$band infrared spectra (1.51 -- 1.70\,$\mu$m) at a resolving power 
$R\sim$22 500 \citep{Majewski+2010}. Stellar atmospheric parameters ($T_{\rm eff}$, log\,$g$, [M/H]), 
abundance ratios [C/M], [N/M] and [$\alpha$/M], as well as elemental abundances [X/H] for 15 
individual elements are deduced from the spectra with the Apogee Stellar Parameter and Chemical 
Abundance Pipeline (ASPCAP) via template matching with a synthetic spectral 
library. For determinations of $T_{\rm eff}$, log\,$g$, [M/H], [C/M], [N/M] and [$\alpha$/M], the full 
APOGEE spectra are used for template matching, while for determinations of the individual elemental abundances,  
specific segments of spectra are selected for template matching \citep{Garcia_Perez+2015, Holtzman+2015}.
Systematic trends in the resultant metallicity [M/H] and elemental abundances [X/H] (except for carbon abundance [C/H] and nitrogen 
abundance [N/H]) as a function of $T_{\rm eff}$ are corrected for using member stars of open and 
globular clusters, yielding a final internal accuracy of 0.05 -- 0.10\,dex. For some elements, 
the internal accuracy of abundance is even better than 0.05\,dex at high spectral SNR \citep{Holtzman+2015}. 
Values of [M/H] have also been calibrated externally to account for the observed systematic trends with the 
metallicities [Fe/H] of star clusters. Though no external calibration of individual elemental abundances 
against independent data set have been carried out, the overall trend of [$\alpha$/M] with [M/H], as well as the overall 
trends of individual elemental abundances with [Fe/H], seem to match well those deduced 
from high resolution spectroscopy of stars in the solar neighborhood \citep[e.g.][]{Bensby+2003}. 
The SDSS DR12 catalog includes 163\,278 APOGEE stars in total, and 102\,178 of them 
are giant stars with calibrated stellar parameters \citep{Holtzman+2015}. 

\begin{figure}
\centering
\includegraphics[width=80mm]{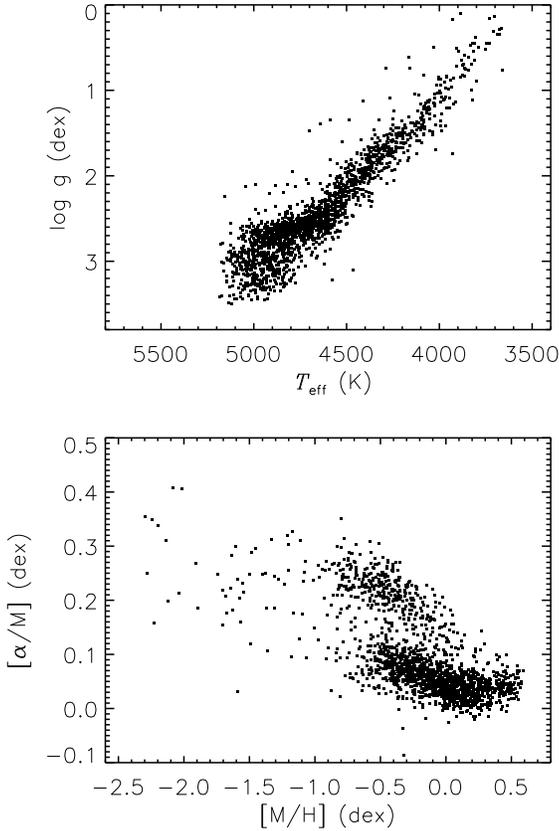}
\caption{Distribution of the LAMOST-APOGEE training stars in the $T_{\rm eff}$ -- log\,$g$ 
and [M/H] -- [$\alpha$/M] planes. Values of $T_{\rm eff}$ are those derived with LSP3, while 
those of log\,$g$, [M/H] and [$\alpha$/M] are from the APOGEE catalog \citep{Holtzman+2015}.}
\label{Fig7}
\end{figure}
A cross-identification of the LSS-GAC catalog with APOGEE giant stars of $3500<T_{\rm eff}<5300$\,K  
and log\,$g < 3.8$\,dex yields 8400 common objects that have a LAMOST spectral SNR 
higher than 10. Here $T_{\rm eff}$ and log\,$g$ refer to the 
APOGEE values. To derive principal components accurately, 
we include only stars with a spectral SNR higher than 50 in the training set. 
This yields a total of  3533 stars in the set. We divide the stars into small cells in 
the [M/H] -- [$\alpha$/M] plane, and select half of them (1766) uniformly and randomly 
from the individual cells as our training sample.  
The other half, as well as common objects of a LAMOST spectral SNR lower than 50, are selected 
as the test set. Fig.\,7 shows the distribution of the training stars in the  $T_{\rm eff}$ -- log\,$g$ 
and the [M/H] -- [$\alpha$/M] planes. 
With this training set, we estimate the metallicity [M/H], $\alpha$-element to metallicity ratio [$\alpha$/M], 
$\alpha$-element to iron abundance ratio [$\alpha$/Fe] and elemental abundances [C/H], [N/H] and [Fe/H].  
An investigation on deriving abundances of other individual elements from LAMOST spectra will be presented elsewhere. 
Here we also do not estimate effective temperature and surface gravity, because the 
current LSP3 can provide effective temperature with good accuracy, and we have adopted 
the LAMOST-$Kepler$ stars with asteroseismic measurements of surface gravity as 
the training data set for the estimation of surface gravity for LAMOST giant stars.    

\begin{figure*}
\centering
\includegraphics[width=140mm]{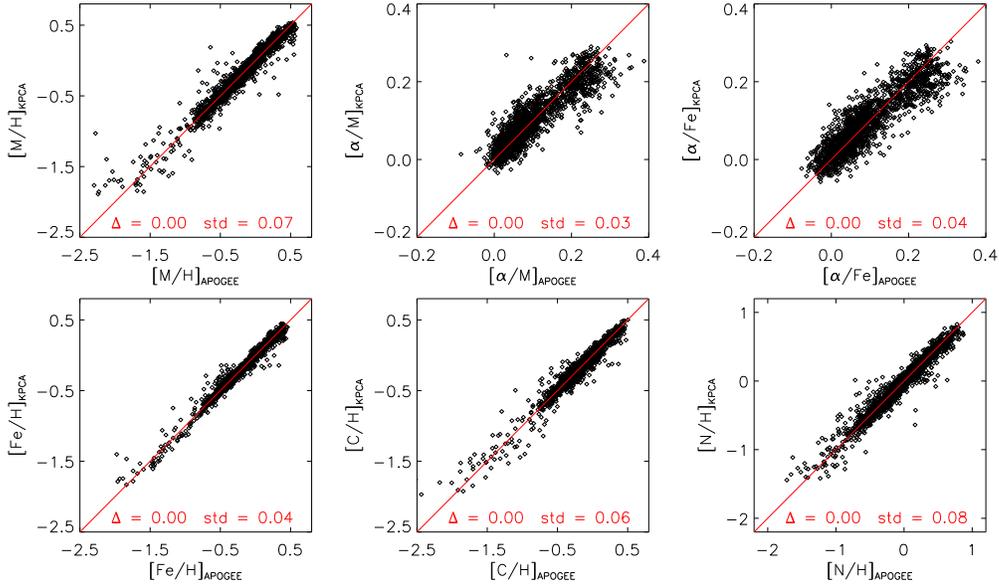}
\caption{Comparison of abundances between the regressed values and the APOGEE measurements 
for the LAMOST-APOGEE training stars.  
Resistant estimates of the mean and standard deviation of the residuals are marked in the plots.}
\label{Fig8}
\end{figure*}
Similar to the case of the LAMOST-$Kepler$ training set, we adopt a number of 90 PCs for 
the multiple-linear regression. Fig.\,8 plots a comparison of the regressed parameters with the 
APOGEE values for the training stars. The Figure shows that the APOGEE parameters are all well 
reproduced by the regression, and the standard deviations of the residuals are only 0.07\,dex 
for [M/H], 0.03\,dex for [$\alpha$/M], 0.04\,dex for [$\alpha$/Fe] and 0.04 -- 0.08\,dex for 
[Fe/H], [C/H] and [N/H]. 
The small standard deviations not only demonstrate the feasibility of our method, but also 
validate the precisions of the APOGEE parameters. 
Values of standard deviation for the metal-poor stars (e.g. ${\rm [M/H]}<1.0$\,dex) are 
larger than those for the metal-rich stars, this is because the APOGEE parameters are less accurate 
for the metal-poor stars. Since there are few training stars of ${\rm [M/H]}<-2.0$\,dex, and the 
parameters of those stars are also quite uncertain, our method probably systematically 
overestimates metallicity and elemental abundances for stars of metallicity below $-2.0$\,dex.          
Similarly, our method probably underestimates [$\alpha$/M] and [$\alpha$/Fe] 
for stars of ${\rm [\alpha/M]} > 0.3$\,dex. Note that since abundances of the APOGEE 
measurements are not externally calibrated to standard absolute scales, systematic offsets or 
bias could be hided in the APOGEE abundances, which may also have been propagated into our results.  

\begin{figure*}
\centering
\includegraphics[width=180mm]{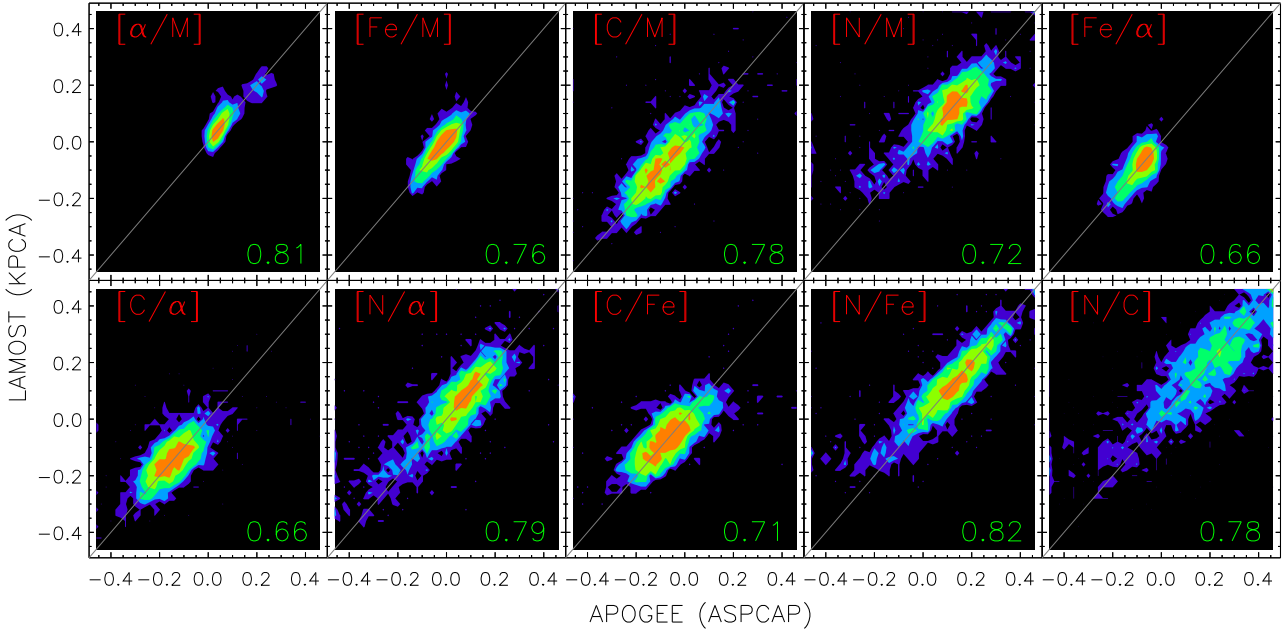}
\caption{Comparison of elemental abundance ratios between the  
KPCA and APOGEE results for the LAMOST-APOGEE test star sample. 
Colours indicate stellar number density in logarithmic scale. The number in each plot labels 
linear correlation coefficient calculated with the individual stars.}
\label{Fig9}
\end{figure*}

Although the standard deviations of regression residuals are small for all abundances, 
it not necessarily means that all the abundances are direct and reliable estimates, 
which reflect abundances of $the$ individual elements. This is because our method is 
deducing abundance using all spectral features, not only those of $the$ element 
concerned. As a consequence, it is possible that abundances of some elements are 
tied to those of other elements artificially, if the former contribute too weak features in the spectra.
To examine such artifacts, we further compare abundances ratios [X/Y] deduced from our 
abundance estimates with those of the APOGEE values for the test star sample. Here 
X (and Y) refers to any of the metal M, alpha-elements $\alpha$, iron Fe, carbon C and nitrogen N, 
and the $\alpha$ abundance is transferred from [$\alpha$/Fe] and [Fe/H].   
For the comparison, we select stars 
with a LAMOST spectral SNR higher than 50 and with a APOGEE spectral SNR higher than 60, 
and we further discard stars with uncertainties of the APOGEE abundances larger than 0.1\,dex. 
The comparisons, as well as correlation coefficients between our abundance ratio estimates 
and those of the APOGEE, are shown in Fig.\,9. The Figure shows that for all abundance   
ratios, our results are in good agreement with the APOGEE values, indicating our abundances are 
reliable estimates. We also inspect the LAMOST spectra in given limited ranges of 
$T_{\rm eff}$, log\,$g$, and [Fe/H], and find that strength of spectral features of C, N and $\alpha$-elements 
(e.g. CN $\lambda$4215{\AA}, CH $\lambda$4314{\AA}, Ca\,{\sc i} $\lambda$4226{\AA} etc.) 
are correlated well with the estimated abundances.

\section{results}
By June, 2014, LSS-GAC has collected more than 1.8 million LSS-GAC spectra with a spectral SNR 
higher than 10 for about 1.4 million unique stars. Fundamental stellar atmospheric parameters 
($T_{\rm eff}$, log\,$g$, [Fe/H]) have been deduced from these spectra with LSP3 (v1) 
via a template-matching technique utilizing the MILES spectral library. 
Here we derive stellar parameters from these spectra with the KPCA-based regression.      
Note that the LSS-GAC spectra used in this work are processed at Peking University 
with the LAMOST 2D pipeline \citep[v2.6;][]{Luo+2015} and the flux calibration pipeline 
developed specifically for the LSS-GAC \citep{Xiang+2015b}.

\subsection{Examining the method with duplicate observations} 
\begin{figure}
\centering
\includegraphics[width=80mm]{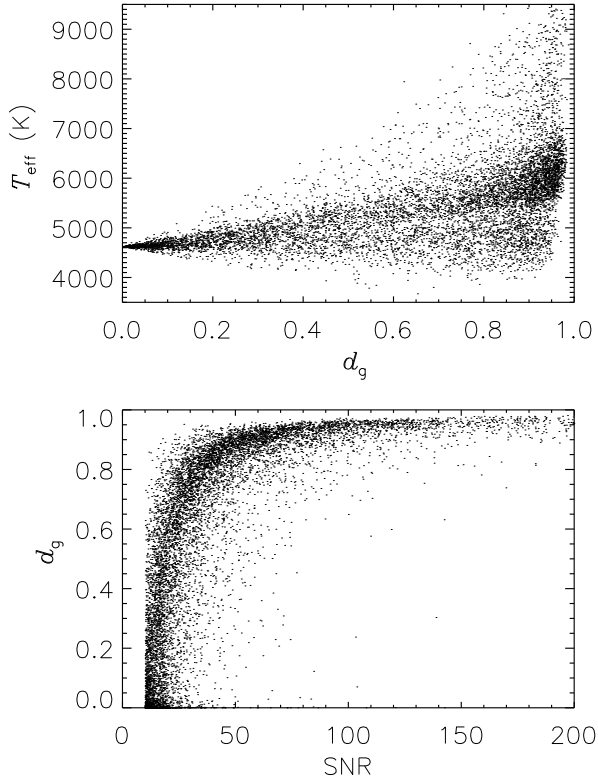}
\caption{$Upper$ $panel$: The estimated effective temperatures as a function of $d_{\rm g}$ (see the text $\S{5.1}$). 
$Lower$ $panel$: The $d_{\rm g}$ values as a function of the spectral signal-to-noise ratio. 
Only results of 10\,000 stars randomly selected from the whole sample are shown in the plots.}
\label{Fig10}
\end{figure}
For a given target spectrum $\bm{x}_{\rm target}$, let $d_{\rm g}$ denote the maximal value of the kernel function,  
\[d_{\rm g} \equiv k(\bm{x}_{\rm target},\bm{x}_{\rm train}) = {\rm exp}(\frac{-\|\bm{x}_{\rm target}-\bm{x}_{\rm train}\|^2}{c}).\] 
Here $\bm{x}_{\rm train}$ represents the training spectrum and the subscript g means `Gaussian'.  
The $d_{\rm g}$ is a metric describing similarities between the target and training spectra. 
If a target spectrum is exactly the same as one of the training spectra, then $d_{\rm g}$ 
is unity. A small value of $d_{\rm g}$ could be due to either poor quality (mostly low SNR) 
of the target spectrum or the fact that the target spectrum is so special that none of the training spectra match it. 
If the analysis yields a small value of $d_{\rm g}$, one expects that the estimated atmospheric 
parameters are of low accuracy. As an example, Fig.\,10 plots the estimated $T_{\rm eff}$ 
for the LSS-GAC stars using the MILES as the training set as a function of $d_{\rm g}$. 
The Figure shows that as $d_{\rm g}$ diminishes to 0, the estimated $T_{\rm eff}$ converages to a 
specific value around 4600\,K. Similar artifacts are also found in the case of log\,$g$ and [Fe/H] estimation. 
The artifacts arise mainly from defects in the spectra, especially at low spectral SNRs. 
The bottom panel of Fig.\,10 illustrates that the $d_{\rm g}$ value is very sensitive to spectral SNR. 
A small $d_{\rm g}$ value (e.g. $< 0.5$) is most likely from spectra of SNR lower than 20. 
Since there are a large number of spectra of low SNRs in the LAMOST archive, it is very important to 
overcome those artifacts in order to avoid dramatic systematic bias in the estimated parameters. 

\begin{figure}
\centering
\includegraphics[width=80mm]{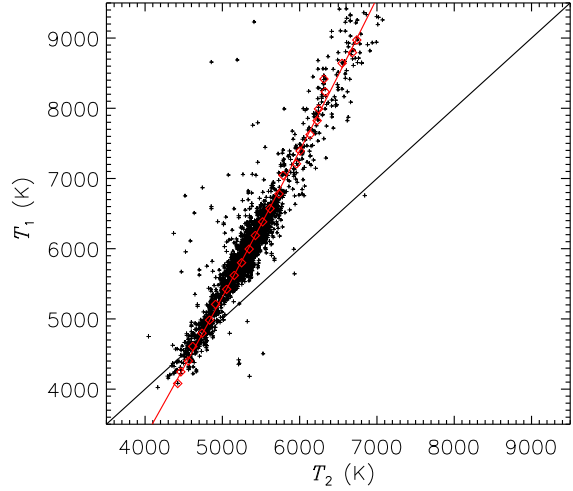}
\caption{Comparison of effective temperatures deduced from spectra that yield $0.5<d_{\rm g}<0.55$ ($T_2$)
with their counterparts derived from duplicate observations of $d_{\rm g}>0.9$ ($T_1$). 
The red diamonds are mean values of $T_2$ in different bins of $T_1$, and the red line 
is a 2nd-order polynomial fit to the red diamonds.}
\label{Fig11}
\end{figure}
Fortunately,  in our sample there are a substantial number of stars with duplicate observations 
that yield different values of $d_{\rm g}$. This allows us to carry out an internal calibration to 
correct for the artifacts for small values of $d_{\rm g}$. In doing so, 
we search for stars with duplicate spectra 
by requiring that one yields $d_{\rm g} > 0.9$ whereas 
the other has $d_{\rm g} < 0.9$. In total, 143,788 pairs of duplicate spectra are found. 
Parameters deduced from spectra that yield $d_{\rm g} > 0.9$ are supposed 
to be free from the artifacts. Results derived from spectra that yield $d_{\rm g} < 0.9$ are 
grouped into bins of $d_{\rm g}$ with a bin size of 0.05. For each bin, parameters derived from 
the duplicate spectra sets are fitted by polynomials, and the resultant polynomials 
are used to calibrate parameter values derived from spectra of small $d_{\rm g}$ to those 
deduced from spectra of $d_{\rm g} > 0.9$. 
Here we adopt a 2nd-order polynomial for the calibration of effective temperature, and 
linear function for the calibration of other parameters.   
As an example, Fig.\,11 plots the polynomial fit of $T_{\rm eff}$ deduced from the 
duplicate spectra sets for the case that the lower quality set have $0.5<d_{\rm g}<0.55$. 
The Figure shows that a 2nd-order polynomial fit of $T_1$ as a function of $T_2$ is sufficient. 
Here $T_1$ represents $T_{\rm eff}$ derived from the set of spectra
with $d_{\rm g} > 0.9$, and $T_2$ represents those derived from the other set of 
spectra with $0.5<d_{\rm g} < 0.55$. 
Note that Fig.\,11 also shows a few outliers. 
For instance, there are several stars with $T_1 > 7500$\,K and $T_2<6000$\,K. 
For those stars, LSP3 parameters deduced from the duplicate spectra also exhibit significant scatter. 
Further inspection shows that the duplicate spectra of those stars  
do vary significantly, indicating that they are indeed variables. 
  
In the following subsections, we will not repeat the introduction of 
the corrections for systematic bias of parameters for low values of $d_{\rm g}$. 
All results presented refer to those after the internal calibrations. 
In addition, parameters deduced from spectra 
with $d_{\rm g} < 0.2$ are discarded due to potentially large uncertainties 
of the internal calibrations.

\subsection{Stellar atmospheric parameters derived with the MILES library}

As introduced in \S{4.1}, we estimate KPCA effective temperature and metallicity for all 
stars of $T_{\rm eff}^{\rm LSP3}<12000$\,K and surface gravity for stars of
log\,$g_{\rm LSP3} > 3.0$\,dex utilizing the MILES training stars. 
The resultant parameters are examined detailedly by 
comparing results from duplicate observations and comparing with external 
data sets.

\subsubsection{Random errors induced by imperfections of the spectra}

\begin{figure}
\centering
\includegraphics[width=80mm]{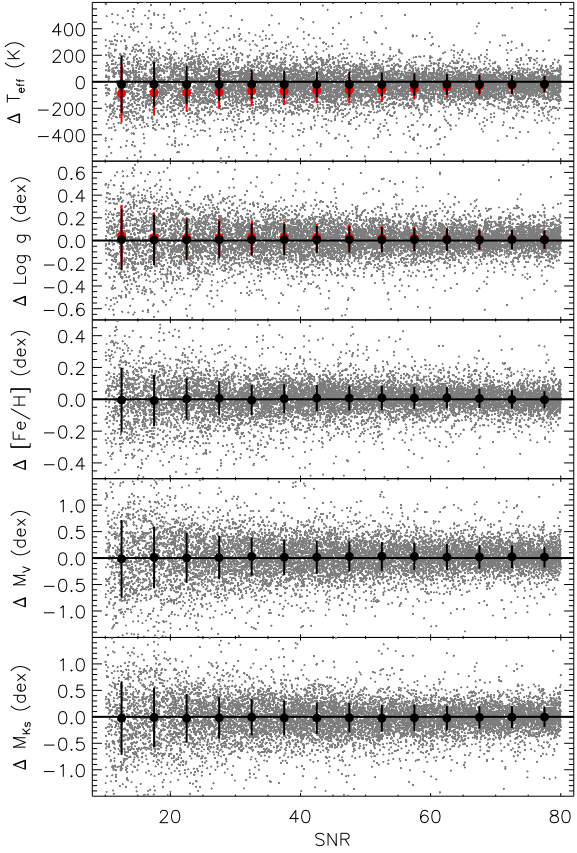}
\caption{Differences of atmospheric parameters and absolute magnitudes derived from duplicate 
observations of different spectral SNR, plotted against the SNR of lower-quality observation. 
The SNR of the higher quality spectrum is required to be larger than 80 per pixel. 
Parameter values derived from the lower-quality set of duplicate observations have been 
internally re-calibrated accounting for small values of $d_{\rm g}$ (see text for detail) 
before the differences are calculated.
The red and black dots are the mean differences before and after a correction 
for the trend with SNR, respectively. 
The error bars are the standard deviations for stars in the individual bins of the SNR}
\label{Fig12}
\end{figure}
Fig.\,12 plots the differences of atmospheric parameters deduced from duplicate observations 
as a function of the lower SNR. As default, systematics in parameters of stars with low $d_{\rm g}$ 
values have been corrected as described above.
Here the duplicate observations refer to those carried out in different nights. 
The Figure shows that $T_{\rm eff}$ deduced from spectra of low SNR 
are slightly underestimated compared to those for high spectral SNR, and 
the underestimation reaches about 100\,K at a SNR of 10. 
Log\,$g$ of stars with SNR $\sim$10 exhibit an overestimation of $\sim$0.05\,dex 
compared to their high SNR counterparts. For [Fe/H], there are no significant 
trends with SNR. We therefore further correct for the systematics in $T_{\rm eff}$ 
and log\,$g$ induced by low spectral SNR via linearly interpolating the mean 
differences as a function of SNR. 
The Figure also shows that the dispersions, which indicate random errors 
of the parameters induced by imperfections of the spectra, are a sensitive function of the spectral SNR, 
and become very small at high SNR. 

\begin{figure*}
\centering
\includegraphics[width=140mm]{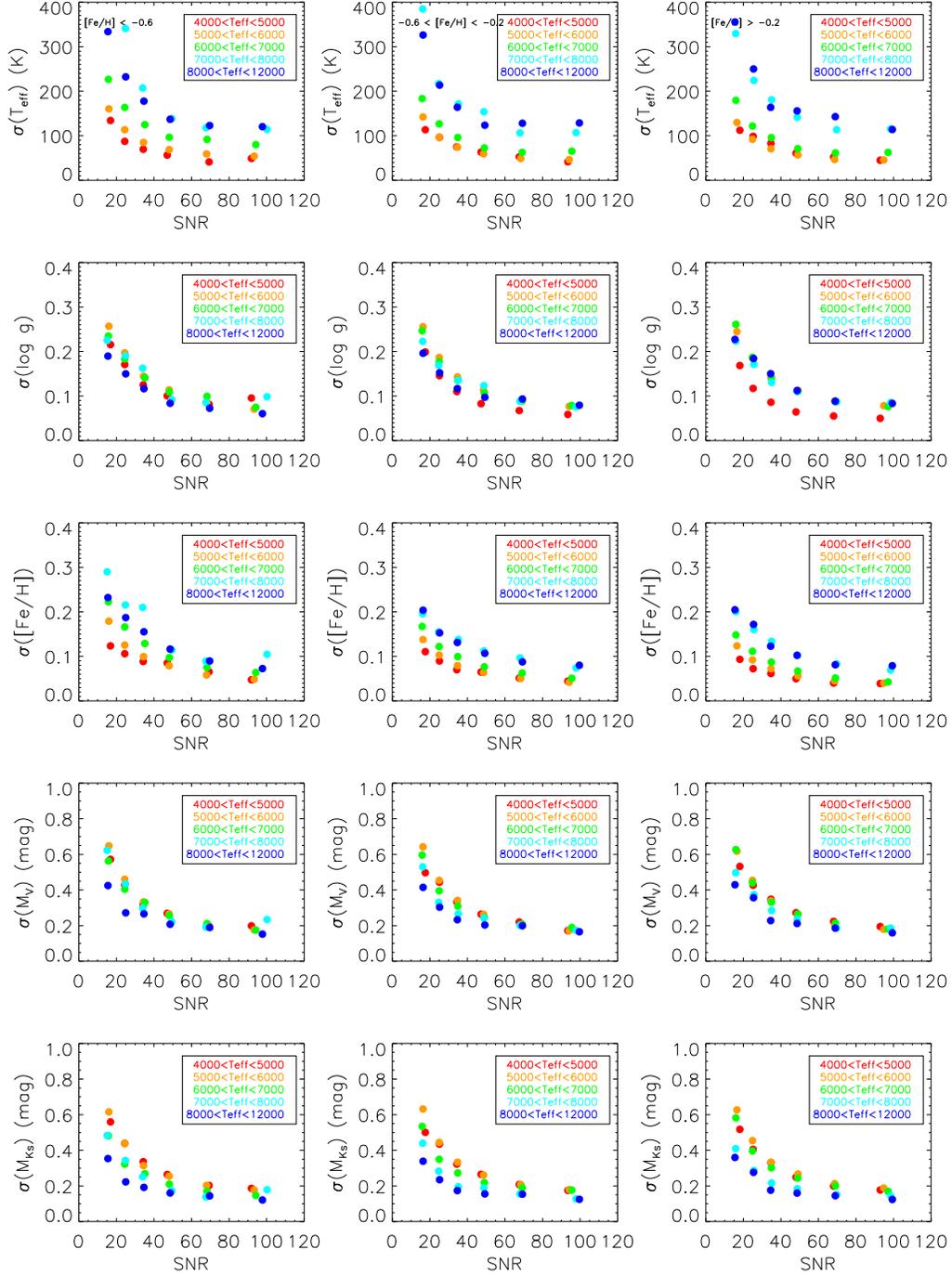}
\caption{Random errors of the estimated parameters, estimated by comparing results 
deduced from duplicate observations of comparable SNRs, as a function of the spectral SNR. 
Stars are grouped into different 
bins of $T_{\rm eff}$ (colour-coded) and [Fe/H] (different columns).}
\label{Fig13}
\end{figure*}
To further investigate random errors of the deduced parameters induced by 
imperfections of the spectra, we select duplicate stars that have comparable 
(within 20 per cent of each other) spectral SNR and are collected in different nights, 
and group them into bins of SNR, $T_{\rm eff}$ and [Fe/H]. In each bin, a resistant 
estimate of the standard deviation of the differences of parameter values derived 
from the duplicate observations is calculated. 
The standard deviations, after divided by the square root of 2, is adopted as the 
random errors induced by imperfections of the spectra. 
Fig.\,13 shows that the random errors are a sensitive function of 
the spectral SNR, and also vary moderately with the spectral type ($T_{\rm eff}$).  
For a SNR of 20, the random error of $T_{\rm eff}$ is 100 -- 120\,K for stars of $T_{\rm eff}<7000$\,K, 
and about 200\,K for stars of $T_{\rm eff}>7000$\,K. For a higher SNR of 50, the random error 
of $T_{\rm eff}$ decreases to $\sim$50\,K for cool stars, and to $\sim$100\,K for those hot ones.  
No significant variations in the random errors are seen for stars of different [Fe/H]. 
For log\,$g$, the random error is about 0.15\,dex for a SNR of 20, and decreases to 0.05\,dex at 
high SNRs, without significant variations with $T_{\rm eff}$ or [Fe/H]. 
The random error of [Fe/H] is close to 0.1\,dex for a SNR of 20 for stars of $T_{\rm eff} < 7000$\,K, 
and decreases to $\sim$0.05\,dex given high SNR. While for hot ($T_{\rm eff}>7000$\,K) 
or metal-poor stars, the random errors amount to 0.15 -- 0.20\,dex for a SNR of 20.  
Note that the above random errors are not realistic errors of the estimated parameters, 
but refers to only errors induced by uncertainties in the observed spectra. The realisitic errors 
should also include those induced by the method of analysis, which can be approximated by the 
regression residuals of the training set, as well as potential systematic bias in the 
MILES stellar atmospheric parameters, which will be discussed in the following subsections.  

\subsubsection{Comparing [Fe/H] with high-resolution spectroscopy}

\begin{figure}
\centering
\includegraphics[width=80mm]{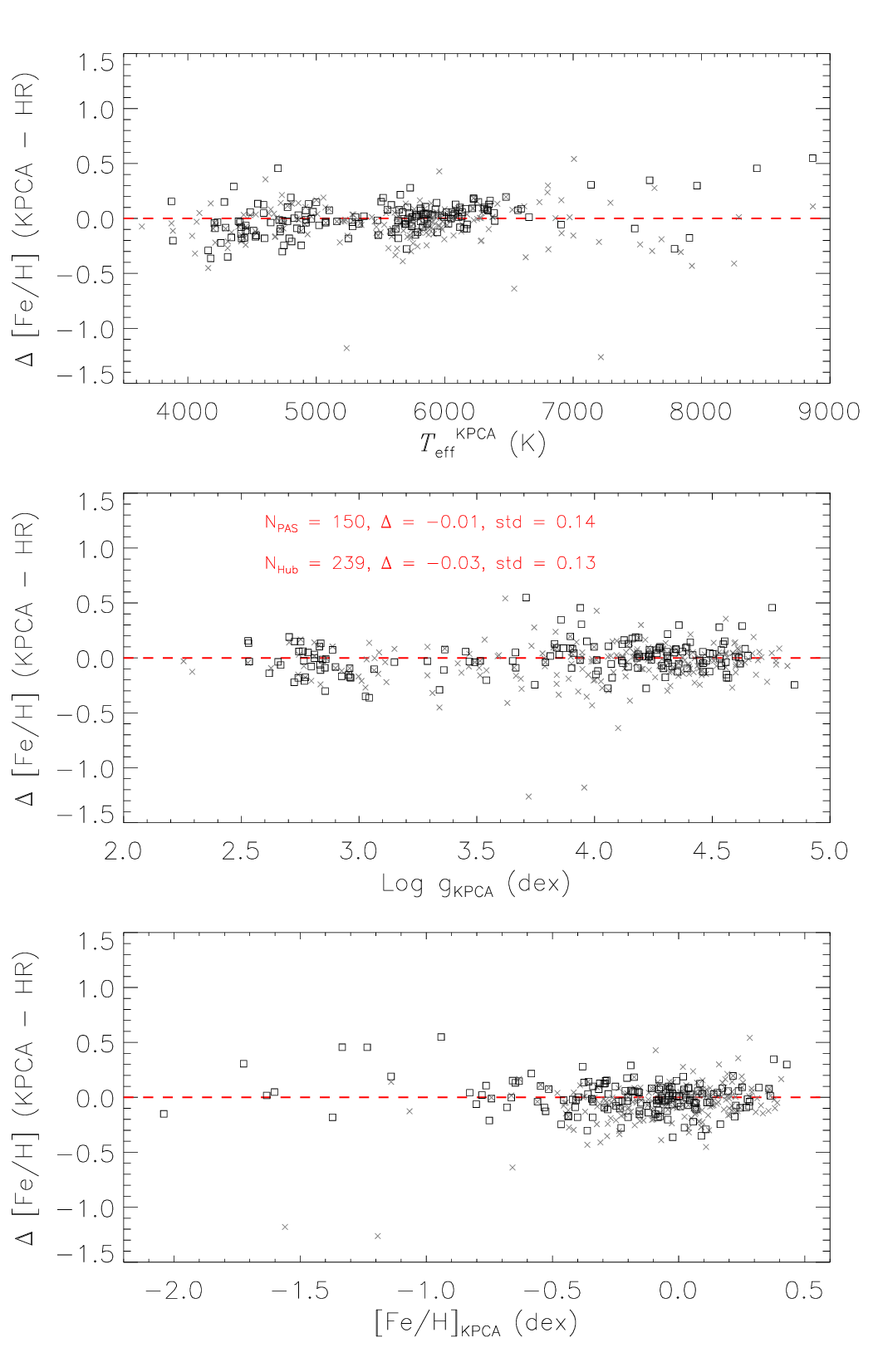}
\caption{Differences between [Fe/H] estimated by KPCA utilizing the MILES training set and 
values of high-resolution spectroscopic measurements, plotted against the KPCA atmospheric parameters. 
Squares and crosses represent high-resolution results from the PASTEL catalog and   
\citet{Huber+2014}, respectively.   
The number of stars in common with the two catalogues,  as well as values from a resistant estimate of the mean 
and standard deviation of the differences, are marked in the middle panel of the plot.}
\label{Fig14}
\end{figure}
We compare our [Fe/H] estimates with two external high-resolution spectroscopic data sets: 
the PASTEL catalog and the catalog of \citet{Huber+2014}. The PASTEL catalog \citep{Soubiran+2010} 
is a collection of literature determinations of stellar atmospheric parameters for 16 649 stars, 
about 6000 of which have [Fe/H] determinations from high-resolution spectroscopy. 
The catalog of \citet{Huber+2014} collects stellar parameters from the recent literatures 
published in recent years for stars in the $Kepler$ field. The catalog contains 819 stars 
whose [Fe/H] are deduced from high-resolution spectroscopy.  
For the comparison, we exclude stars with a LAMOST spectral SNR lower than 20, 
as well as stars whose [Fe/H] in the PASTEL catalog are from literatures published before 1990. 
After the cuts, we have a sample of 150 stars in common with the PASTEL catalog, 
and a sample of 239 stars in common with the catalog of Huber et al. 

Fig.\,14 plots the differences between our [Fe/H] estimates and the high-resolution 
results against $T_{\rm eff}$, log\,$g$ and [Fe/H]. A resistant 
estimate of the mean and standard deviation for the overall sample of   
stars in common with the PASTEL catalog are $-0.01$\,dex and 0.14\,dex, and those values 
are $-0.03$\,dex and 0.13\,dex for stars in common with the catalog of Huber et al.  
The differences are not homogeneously distributed across the parameter space. 
For both samples, the dispersions for stars of $T_{\rm eff} > 7000$\,K are significantly 
larger than those for the bulk, dominated by F/G/K stars of $4000<T_{\rm eff} <7000$\,K.  
In fact, a Gaussian fit to the distribution of the differences yields a dispersion, 
which reflects value for the bulk stars, of 0.10\,dex only for both samples. 
Note that [Fe/H] uncertainties of the high-resolution results themselves are expected 
to be at 0.1\,dex level. No significant systematic trends are seen in the data, 
but there are several outliers showing very large deviations from the high resolution determinations. 
For those outliers, it is found that their literature values of $T_{\rm eff}$ differ 
significantly with our estimates, but our values are consistent with results from 
LSP3. A further inspection of LAMOST spectra for those stars corroborates the 
robustness of our determinations. 
  
\begin{figure}
\centering
\includegraphics[width=80mm]{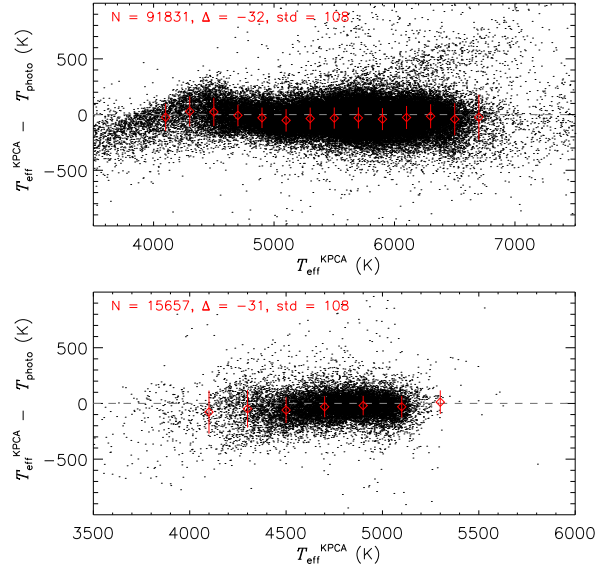}
\caption{Differences between the estimated effective temperature $T_{\rm eff}^{\rm KPCA}$ 
with photometric temperature $T_{\rm photo}$ deduced from the metallicity-dependent 
colour-temperature relation of \citet{Huang+2015b} for dwarf stars (upper) 
and giant stars (lower). The red diamonds and error bars are the mean differences 
and standard deviations at different bins of $T_{\rm eff}^{\rm KPCA}$. The number 
of stars, as well as a resistant estimate of the overall mean and standard deviation of the 
differences, are marked in the plot.}
\label{Fig15}
\end{figure}

\subsubsection{Comparison with photometric temperatures}
We compare our estimates of effective temperature with photometric temperature yielded by  
the metallicity-dependent colour-temperature relation of \citet{Huang+2015b}. 
The relation is derived based on more than a hundred nearby stars with direct effective temperature 
measurements, and are applicable for stars of ${\rm [Fe/H]}>-1.0$\,dex. 
The de-reddened $g-K_{\rm s}$ colour and the above estimated [Fe/H] are used to generate  
photometric temperatures for our sample stars. Here the SDSS $g$-band photometry 
is from the XSTPS-GAC survey \citep{Zhang+2014, Liu+2014} for stars of $g>13.5$, 
and from the APASS survey \citep{Henden+2014} for the brighter stars, 
while the $K_{\rm s}$- band photometry is from the 2MASS catalog \citep{Skrutskie+2006}. 
The $g-K_s$ colour is de-reddened with interstellar reddening $E(B-V)$ from the 
map of \citet[hereafter SFD]{Schlegel+1998} and extinction 
coefficients from \citet{Yuan+2013}. To reduce errors from uncertainties in interstellar 
extinction, we select stars of Galactic latitude $|b|>10\degr$ and $E(B-V)_{\rm SFD} < 0.05$\,mag 
only for the determination of photometric temperatures. 
To ensure high accuracy of the photometric data, we further require the stars 
have a magnitude of $10.5<g<18.5$\,mag, $6<K_{\rm s}<15.5$\,mag, and magnitude errors 
smaller than 0.05\,mag in both bands.
For the comparison, we also require the stars to have a spectral SNR\,$>20$ and ${\rm [Fe/H]}>-1.0$\,dex. 
The above criteria lead to 91\,831 dwarf stars (log\,$g>3.8$\,dex) and 15\,657 giant stars (log\,$g<3.5$\,dex) 
in our sample.  

Fig.\,15 plots the differences between our estimates of effective temperatures $T_{\rm eff}^{\rm KPCA}$ 
and the photometric estimates $T_{\rm photo}$ as a function of the former. The   
mean and standard deviation of the differences are $-32$ and 108\,K, respectively, 
for dwarf stars of $3500<T_{\rm eff}^{\rm KPCA}<7500$\,K. No obvious trend 
of differences is seen in the temperature range 4000 -- 7000\,K. 
Note that the uncertainties of photometric temperatures are believed to be about 
2 per cent \citep{Huang+2015b}. The number of hot
($T_{\rm eff}^{\rm KPCA}> 7000$\,K) stars are too small so that we do not discuss 
those stars here. For stars cooler than 3800\,K, our estimates maybe not 
reliable due to a lack of training stars.  
For giant stars, the differences have a mean and standard deviation of $-31$ and 108\,K, 
respectively. Again, no significant trend of differences with temperature is seen, 
except for stars with extremely temperatures ($T_{\rm eff}^{\rm KPCA} < 4500$\,K), 
foe which our estimates seem to be lower than the photometric estimates by a few tens Kelvin.

\subsubsection{Uncertainties in the surface gravity}
In Fig.\,1, we show that given a PC number of 100, our estimates 
of log\,$g$ from the LAMOST spectra are consistent well with the asteroseismic measurements 
for the LAMOST-$Kepler$ common stars, with a mean difference and standard deviation 
of only 0.00\,dex and 0.11\,dex, respectively, suggesting that our log\,$g$ estimates 
have achieved a precision of 0.1\,dex. However, this is only realistic for 
stars with high spectral SNRs because most of the LAMOST-$Kepler$ stars have a 
spectral SNR higher than 50.
Considering the random errors of log\,$g$ are sensitive to SNR as illustrated by Fig.\,13, 
we expect that uncertainties of the log\,$g$ estimates increase to 0.2\,dex for a SNR of 20.    
Note that here the asteroseismic sample contains only stars of 
$5500<T_{\rm eff}^{\rm KPCA}<6500$\,K, thus log\,$g$ estimates for stars 
of temperatures outside this range need to be further examined. It is expected 
that log\,$g$ estimates for stars of higher temperatures have larger 
uncertainties because, as Fig.\,2 demonstrates, the method errors become significantly larger at higher temperatures.  
 
\begin{figure*}
\centering
\includegraphics[width=160mm]{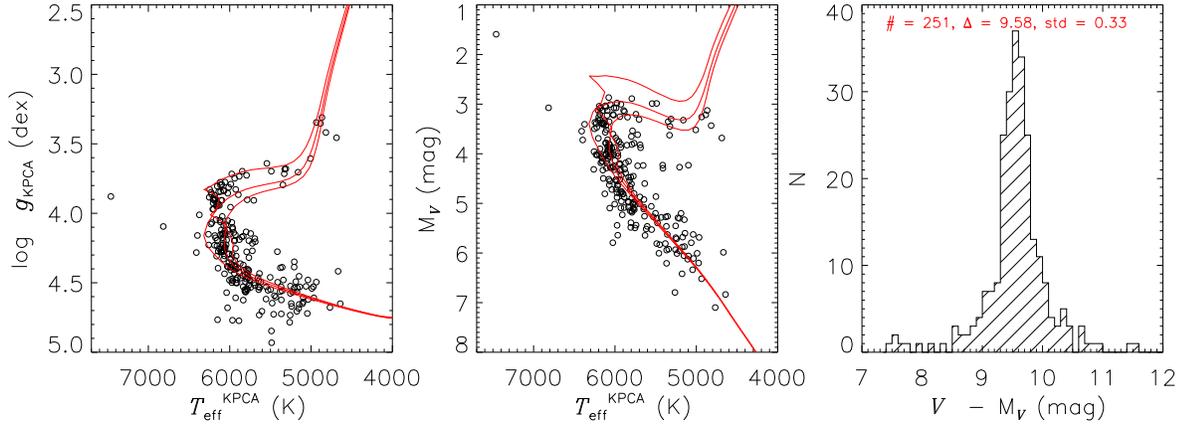}
\caption{Distribution of member star candidates of M\,67 in the $T_{\rm eff}$ -- log\,$g$ 
(left) and the $T_{\rm eff}$ -- ${\rm M}_V$ diagrams. The red lines are Yonsei-Yale isochrones 
with solar metallicity and ages of 3, 4 and 5\,Gyr from left to right, respectively. The right panel plots the 
histogram of distance modulus in $V$-band derived from the individual member candidates. 
The number of stars, mean and standard deviation of the distribution are marked.}
\label{Fig16}
\end{figure*}  
We make a further sanity check of log\,$g$ estimates with member star candidates of 
open cluster M\,67, which has a literature metallicity of $-0.01$\,dex \citep{Jacobson+2011} 
and an age of 4.3\,Gyr \citep{Richer+1998, Salaris+2004}. 
Specific observations are designed to target member candidates of 
open clusters utilizing test observation nights of the LAMOST (Yang et al., in preparation), 
and we select the cluster member stars based on the spatial position, the Hertzsprung-Russel 
(HR) diagram and the radial velocity yielded by the LSP3 in the same way as \citet{Xiang+2015a}. 
In total, more than 500 member star candidates of M\,67 are selected. Here for the sanity 
check we pick out stars with a SNR higher than 20 and a log\,$g$ yielded by the LSP3 
template matching method larger than 3.0\,dex. These criteria lead to 251 unique member 
candidates, whose distribution in the $T_{\rm eff}$ -- log\,$g$ plane are plotted in the left 
panel of Fig.\,16. Also plotted in the Figure are three Yonsei-Yale (Y$^2$) isochrones \citep{Demarque+2004} 
with solar metallicity and ages of 3, 4 and 5 Gyr, respectively. The Figure illustrates that 
though there are some outliers, the main sequence and the main sequence turn-off stars 
are consistent well with the isochrones of 4\,Gyr, while log\,$g$ of the bulk of subgiant stars 
are lower than the isochrone values by about 0.1\,dex.

\subsection{Absolute magnitudes estimated with the LAMOST-Hipparcos training set} 
Absolute magnitudes in $V$ and $K_{\rm s}$ bands for all LSS-GAC stars of 
$T_{\rm eff}^{\rm LSP3} < 12000$\,K are derived with the LAMOST-Hipparcos training data set. 
Detailed internal examinations and calibrations are first carried out as introduced in \S{5.1}. 

The bottom panels of Fig.\,12 illustrate that though low spectral SNR induces larger  
random errors, it does not cause systematic bias to our results. 
A detailed investigation of random errors induced by spectral imperfections 
for stars of different SNRs and stellar atmospheric parameters is presented in Fig.\,13.  
The Figure shows that the random error is a steep function of spectral SNR, and 
also vary with spectra types. At a SNR of 20, random errors of both ${\rm M}_V$ 
and ${\rm M}_{K{\rm s}}$ vary from 0.3 to 0.5\,mag, depending on effective temperatures 
of the stars, while at a SNR of 50, the random errors decrease to 0.15 -- 0.3\,mag, 
and the values continuously decrease to 0.1 -- 0.15\,dex at high SNRs.  

The middle panel of Fig.\,16 shows that for member star candidates of M\,67, 
the $T_{\rm eff}$ -- ${\rm M}_V$ diagram match well with the Y$^2$ isochrones. 
Not only the main sequence and main sequence turn-off stars, but also the subgiant stars, 
are all consistent with the isochrone of 4\,Gyr. The right panel of the Figure plots 
the distribution of the $V$-band distance modulus of M\,67 derived with the individual stars. 
The distribution yields a mean modulus of 9.58\,mag and a standard deviation of 
0.33\,mag. Outliers are clearly visible in the Figure. Some of the outliers are 
expected to be binaries/multiple stars and/or variable stars. It is also possible 
that some of the outliers are contaminations of field stars. 
In fact, a Gaussian fit, which is less affected by the outliers, yields a dispersion of 
only 0.24\,mag, corresponding to a distance error of 12 per cent. 

\begin{figure}
\centering
\includegraphics[width=80mm]{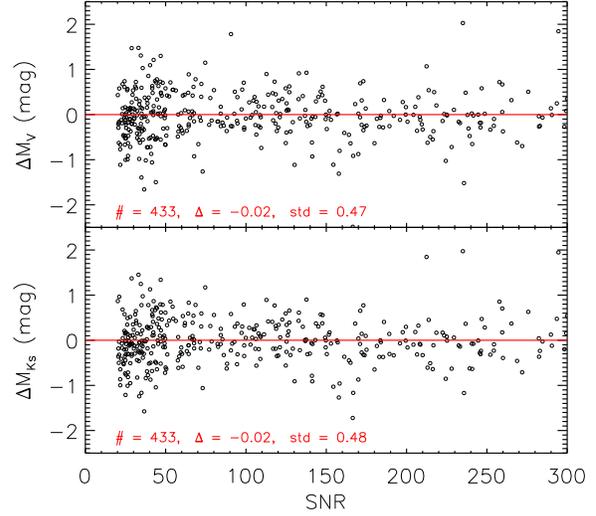}
\caption{Differences of absolute magnitudes between with the estimated values 
and those deduced from the Hipparcos distance for the LAMOST-Hipparcos test stars 
as a function of spectral SNR. The number of stars, the mean and dispersion of 
a Gaussian fit on the overall distribution of the differences are marked in the plot.}
\label{Fig17}
\end{figure}
We further utilize LAMOST-Hipparcos stars that have reliable parallax measurements 
but are not included in the training sample as a test data set to examine our results. 
Those stars are not included in the training data set due to either low spectral SNR ($<50$) 
or that they are duplicate observation counterparts of the training stars. 
Here we select only stars with uncertainties in absolute magnitudes smaller than 0.3\,mag for ${\rm M}_V$ and 
${\rm M}_{K{\rm s}}$ deduced utilizing the Hipparcos parallax, and further require 
that the stars have a spectral SNR higher than 20. 
These criteria lead to 433 stars in our test sample. 
Fig.\,17 plots the differences of absolute magnitudes between our estimates 
and those deduced utilizing the Hipparcos parallax. The differences are plotted against spectral SNR. 
The mean and standard deviation of the overall differences are respectively $-0.02$\,mag 
and 0.47\,mag for ${\rm M}_V$, $-0.02$ and 0.48\,mag for ${\rm M}_{K{\rm s}}$. 
Considering that typical uncertainties of absolute magnitudes propagated from errors 
of the Hipparcos parallax of our sample stars are 0.2 -- 0.3\,mag,  
typical errors of our magnitude estimates should be 0.36 -- 0.43\,mag 
for both ${\rm M}_V$ and ${\rm M}_{K{\rm s}}$. Note however that, uncertainties 
in the estimated absolute magnitudes are sensitive to spectral SNR and spectral types. 
For stars of SNR higher than 50, the above standard deviation becomes 0.4\,mag only, 
corresponding to an uncertainty of 0.3 -- 0.4\,mag in our estimates. 
The Figure also shows a considerable fraction of stars with large differences (e.g. $> 1.0$\,mag). 
By referring to the SIMBAD database \citep{Wenger+2000}, many of those stars 
are binary/multiple or variable stars, while the others are likely caused by spectral 
imperfections --- especially because the LAMOST-Hipparcos stars are too bright 
to be feasible for observation with the LAMOST. In fact, most of the LAMOST-Hipparcos 
stars are observed in conditions with bright lunar light, and for many of them, 
offsets in coordinates are set for fiber positioning to avoid saturation. 
As a result, for those stars, even though the reported SNRs are high, 
their real spectral quality are actually not good, leading to inaccurate parameter estimates.

\subsection{Surface gravities estimated with the LAMOST-$Kepler$ training set} 

 \begin{figure}
\centering
\includegraphics[width=80mm]{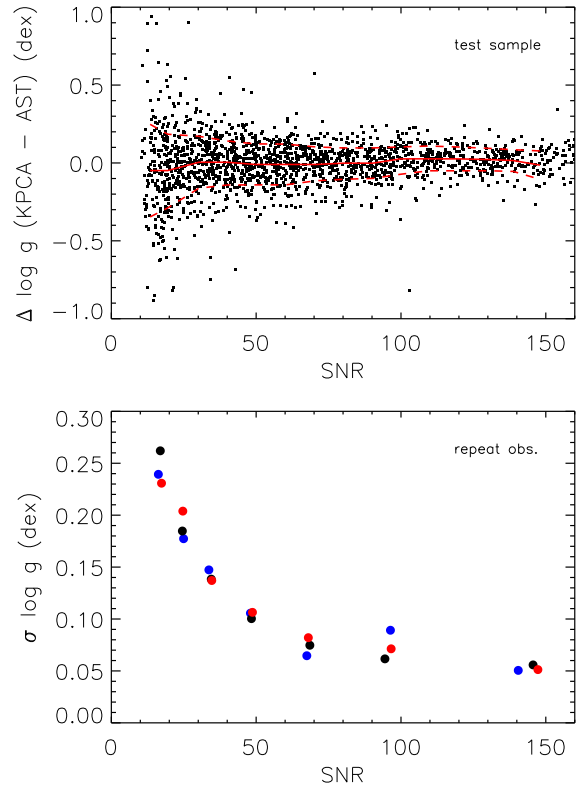}
\caption{$Upper$ $panel$: Differences of the estimated log\,$g$ with the asteroseismic values 
for the LAMOST-$Kepler$ test star sample; $Lower$ $panel$: Random errors of log\,$g$ 
deduced by comparing results of duplicate observations, as a function of the SNR. Different 
colours represent results for stars in different ranges of [Fe/H] (blue: [$-3.0$, $-0.6$]; 
black: [$-0.6$, $-0.2$], red: [$-0.2$, 0.5]).}
\label{Fig18}
\end{figure}

Surface gravity log\,$g$ for all stars of log\,$g_{\rm LSP3} <  3.8$\,dex and 
$T_{\rm eff}^{\rm LSP3} < 5500$\,K are estimated with the KPCA regression 
method using the LAMOST-$Kepler$ training set. Internal calibrations for 
stars of low $d_{\rm g}$ values are carried out as introduced in \S{5.1}. 
As an evaluation of log\,$g$ estimates, the upper panel of Fig.\,18 plots 
the differences between the KPCA log\,$g$ and the asteroseismic values 
for the LAMOST-$Kepler$ test sample (cf. \S{4.2}) as a function of the spectral SNR. 
It shows that the mean differences are close to zero at different SNRs, and the dispersions 
(standard deviations) decrease from $\sim$0.3\,dex at a SNR of 10 to $\sim$0.2\,dex at a SNR of 30, 
and $\sim$0.1\,dex at high SNRs. The results are consistent with those for the training sample, 
which is $-0.01\pm0.11$\,dex (\S{4.2}), considering that the training sample includes only stars with 
SNR higher than 50. Such a precision is comparable to that of \citet{Liuchao+2015}, who 
estimate log\,$g$ from the LAMOST spectra using a SVR method, also trained by the LAMOST-$Kepler$ sample stars. 
Note that for stars with low SNRs (e.g. $<20$), we find a slightly larger dispersion of differences
between our log\,$g$ estimates and the asteroseismic values than Liu et al. 
This is because we use the standard deviation to represent the dispersion, 
while Liu et al. adopt the 1$\sigma$ value of the Gaussian distribution.

The lower panel of Fig.\,18 shows the dispersions of log\,$g$ differences deduced from duplicate 
observations as a function of the spectral SNR. Here we have required that for each star, the duplicate observations have 
roughly the same SNRs, and the dispersions have been divided by the square root of 2. The Figure 
shows that the dispersions decrease from $\sim$0.2\,dex at a SNR of 20 to $\sim$\,0.1\,dex at a SNR 
of 50, and further decrease to 0.05\,dex at high SNRs ($>100$). The dependence 
of the dispersions on metallicity is negligible. Note that for the metal-poor stars, though the dispersions 
are small, it is possible that systematic bias dominate the real uncertainties since there are only a few 
stars of ${\rm [Fe/H]} < -1.0$\,dex in our training sample (cf. \S{4.2}).   
Note also that though the dispersions at high SNRs can be as small as 0.05\,dex, the real uncertainties 
at such SNRs are however, larger than the dispersions. This is because the dispersions deduced from 
duplicate observations only account for uncertainties induced by imperfections of the LAMOST spectra 
but not include uncertainties of the method itself.

Since the dispersions deduced from the LAMOST-$Kepler$ test sample are contributed by both 
imperfections of the LAMOST spectra and inadequacy of the method, they are used to estimate 
realistic errors of log\,$g$ estimates. Errors of log\,$g$ are assigned to individual stars based 
on the spectral SNRs. It seems quite clear that uncertainties in the KPCA log\,$g$ 
estimates are only $\sim0.1$\,dex at high SNRs ($\gtrsim80$).

\subsection{Metallicity and elemental abundances estimated with the LAMOST-APOGEE training set} 

Metallicity [M/H], $\alpha$-element to metal abundance ratio [$\alpha$/M], $\alpha$-element to iron 
abundance ratio [$\alpha$/Fe] and elemental abundances [C/H], [N/H] and [Fe/H] are estimated 
for LSS-GAC stars of log\,$g_{\rm LSP3} < 3.8$\,dex and $T_{\rm eff}^{\rm LSP3} < 5500$\,K utilizing the 
LAMOST-APOGEE training stars, and internal calibrations for stars of low $d_{\rm g}$ values are 
carried out as introduced in \S{5.1}. 
    
\begin{figure*}
\centering
\includegraphics[width=140mm]{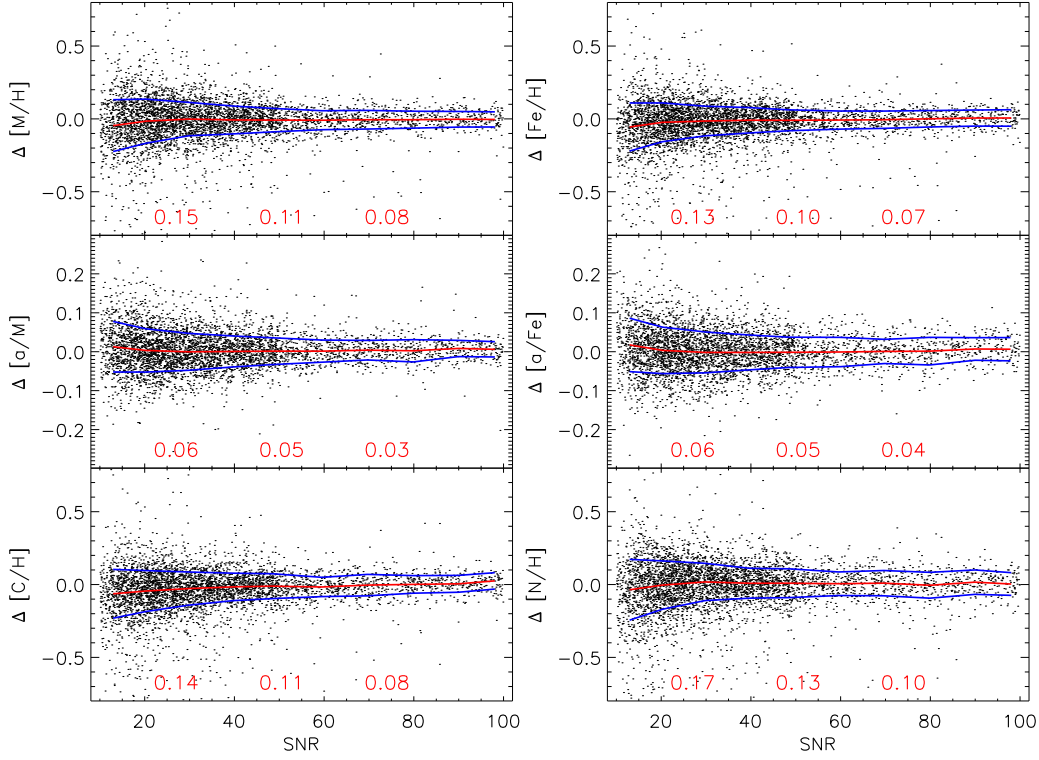}
\caption{Differences of metallicity and elemental abundances between the KPCA estimates 
from the LAMOST spectra and those of the APOGEE measurements for the  
test star sample, as a function of spectral SNR. The mean and standard deviation of the differences as a 
function of SNR are plotted in red and blue solid lines, respectively. In each panel, the three numbers 
labeled are standard deviations calculated at a SNR of 20, 30 and 50, respectively.}  
\label{Fig19}
\end{figure*}

To examine precisions of the estimated abundances, Fig.\,19 plots the differences 
between our estimated values and the APOGEE measurements as a function of LAMOST spectral 
SNR for the test star sample defined in \S{4.4}. The Figure shows that the mean differences at 
all SNR are close to zero for all abundances, and the dispersions decrease significantly with 
increasing SNR at the lower SNR side, while at the higher SNR side, the dispersions keep 
almost flat. Such a trend of dispersions is expected, because for stars with low spectral SNR, 
uncertainties of abundance estimates are dominated by random errors induced by spectral imperfections, 
which is a steep function of SNR, while for stars with high spectral SNR, uncertainties of abundance 
estimates are dominated by method errors, which are mainly propagated from uncertainties 
in the APOGEE measurements of the training stars. 
The three numbers marked in the plot are values of dispersions (standard deviations) calculated 
 at a SNR of 20, 30 and 50, respectively. They show that at a SNR of 20, dispersions of the 
 differences for metal abundance [M/H] and elemental abundance [C/H], [N/H] and [Fe/H]  
 are about 0.13 -- 0.17\,dex. While those numbers decrease to 0.10 -- 0.13\,dex at a SNR of 30, 
 and further decrease to 0.07 -- 0.10\,dex at a SNR of 50. 
 For [$\alpha$/M] ([$\alpha$/Fe]), the dispersions decrease from 0.06\,dex at a SNR 
 of 20 to 0.03\,dex (0.04\,dex) at a SNR of 50. Note that both errors in our abundance estimates 
 and those in the APOGEE measurements have contributed to the dispersions. For stars with high spectral 
 SNR ($>50$), we expect that precision of our estimated metallicity and elemental abundances are 
 comparable to those of the APOGEE measurements, which are 0.05 -- 0.10\,dex for [X/H], and 
 better than 0.05\,dex for [$\alpha$/M] and [$\alpha$/Fe]. 
 Nevertheless, since the APOGEE abundances are determined with $\chi^2$-based 
algorithms and are not externally calibrated to standard scales, potential systematic biases 
could be hided in the APOGEE results, which must have been propagated 
 into our estimates via the training data set. Potential systematic biases are expected 
to be 0.1 -- 0.2\,dex \citep{Holtzman+2015}.

\begin{figure*}
\centering
\includegraphics[width=140mm]{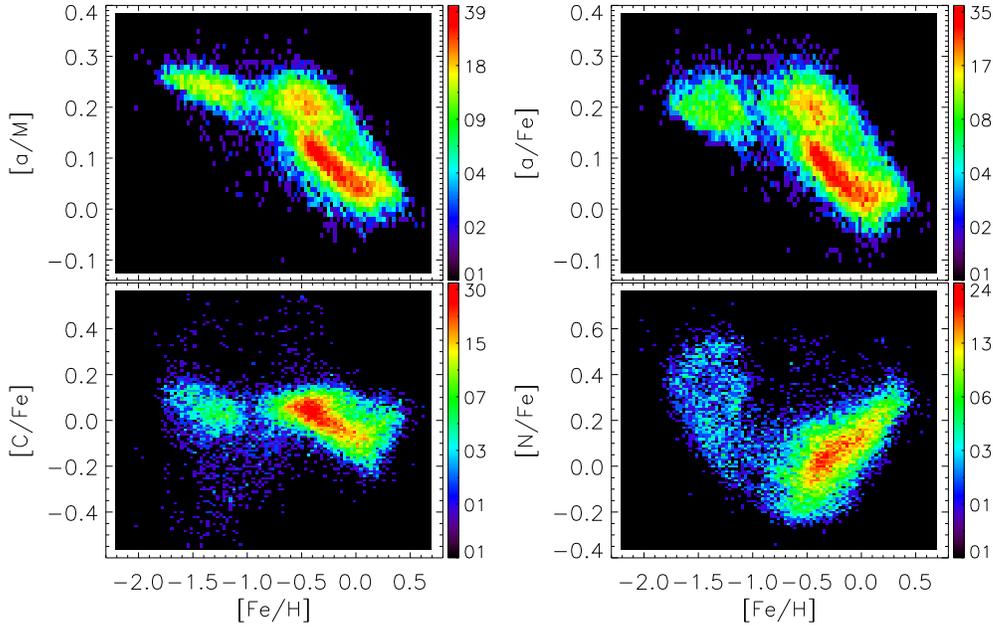}
\caption{Colour-coded stellar number density distributions of LSS-GAC giant stars 
in the [$\alpha$/M] -- [Fe/H] and [X/Fe] -- [Fe/H] planes. Colours indicate 
stellar number density in logarithmic scale. A bin size of 0.03 by 
0.01\,dex is adopted to generate the stellar number density distributions. Only stars of 
Galactic latitudes $|b|>30^\circ$ are shown in the plots.}
\label{Fig20}
\end{figure*}

Fig.\,20 plots the density distribution of LSS-GAC stars with Galactic latitude $|b| > 30^\circ$  
and SNR\,$>30$ in the [X/Fe] (and [$\alpha$/M]) against [Fe/H] plane. For all the elements, morphology of 
the [X/Fe] -- [Fe/H] diagram are resemble to those of the APOGEE stars as shown 
in Fig.\,14 of \citet{Holtzman+2015}.
For the $\alpha$-element to metal abundance ratio [$\alpha$/M], as well as the $\alpha$-element  
to iron abundance ratios [$\alpha$/Fe], there is a decreasing trend with [Fe/H], 
which means that the $\alpha$-element abundances are enhanced for metal-poor stars. 
There seems to be also two bulk of stars in [$\alpha$/M] ($\alpha$/Fe) for a [Fe/H] around $-0.5$\,dex. 
Those trends and features are well consistent with previous results from high resolution spectroscopy
for stars near the solar neighbourhood. The decreasing trend of [$\alpha$/Fe] with [Fe/H] 
is a consequence of Galactic chemical evolution, and the two bulk of stars are likely 
corresponding to the widely investigated thin and thick disk 
star sequences. Note that if we plot all stars in our sample rather than stars of $|b| > 30^\circ$, 
a clear discrimination of the two bulks will not be visible, simply because the thin disk sequence 
dominates our total star sample.
Nevertheless, as described in \S{4.4}, our elemental abundances are probably systematically 
overestimated for very metal-poor stars (${\rm [M/H]}<-2.0$\,dex) due to a lack of metal-poor training 
stars, resulting a cut off of [Fe/H] around $-1.8$\,dex in Fig.\,20, as well as a steeper 
decreasing trend of [N/Fe] with [Fe/H] in the metal-poor side. 

\begin{figure}
\centering
\includegraphics[width=80mm]{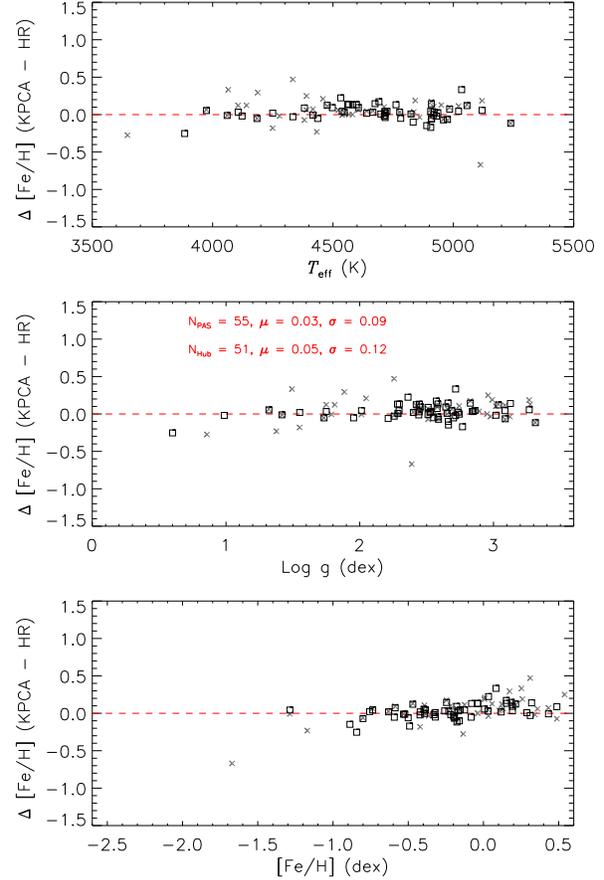}
\caption{Differences between the [Fe/H] estimates utilizing the LAMOST-APOGEE training stars  
and those of high-resolution spectroscopic measurements. The squares represent high-resolution results 
from the PASTEL catalog, while the crosses are from the catalog of \citet{Huber+2014}. 
The differences are shown as a function of the estimated atmospheric parameters.   
The number of stars for each data set,  as well as a resistant estimate of the mean 
and standard deviation, are marked in the middle panel.}
\label{Fig21}
\end{figure}
 Fig.\,21 shows an external comparison of the estimated [M/H] with high-resolution spectroscopic 
 measurements of [Fe/H] from the PASTEL catalog and the catalog of \citet{Huber+2014}. 
 Here we compare the [M/H] but not the [Fe/H] with high resolution spectroscopy because 
 the [M/H] of APOGEE stars have been externally calibrated to [Fe/H] of star clusters, while 
 the [Fe/H] of APOGEE stars are not externally calibrated.  
 In the Figure, we use the tab [Fe/H] to replace [M/H] for consistency. 
 The Figure shows that the overall mean differences are only 0.03
 and 0.05\,dex for the two high-resolution spectroscopy samples, 
 with a standard deviation of 0.09 and 0.12\,dex, respectively. 
 The small positive mean differences are contributed by stars of super-solar metallicities, for 
 which our estimates are systematically higher than the high-resolution spectroscopic 
 data sets. No significant biases with $T_{\rm eff}$ and log\,$g$ are seen. 
  
Finally, according to the comparisons of metallicity and abundances with the APOGEE measurements 
for the test star sample (Fig.\,19), we assign parameter errors to individual stars based on their spectral SNR. 

\subsection{Comparison with the weighted-mean parameters of the LSP3}

\begin{figure*}
\centering
\includegraphics[width=160mm]{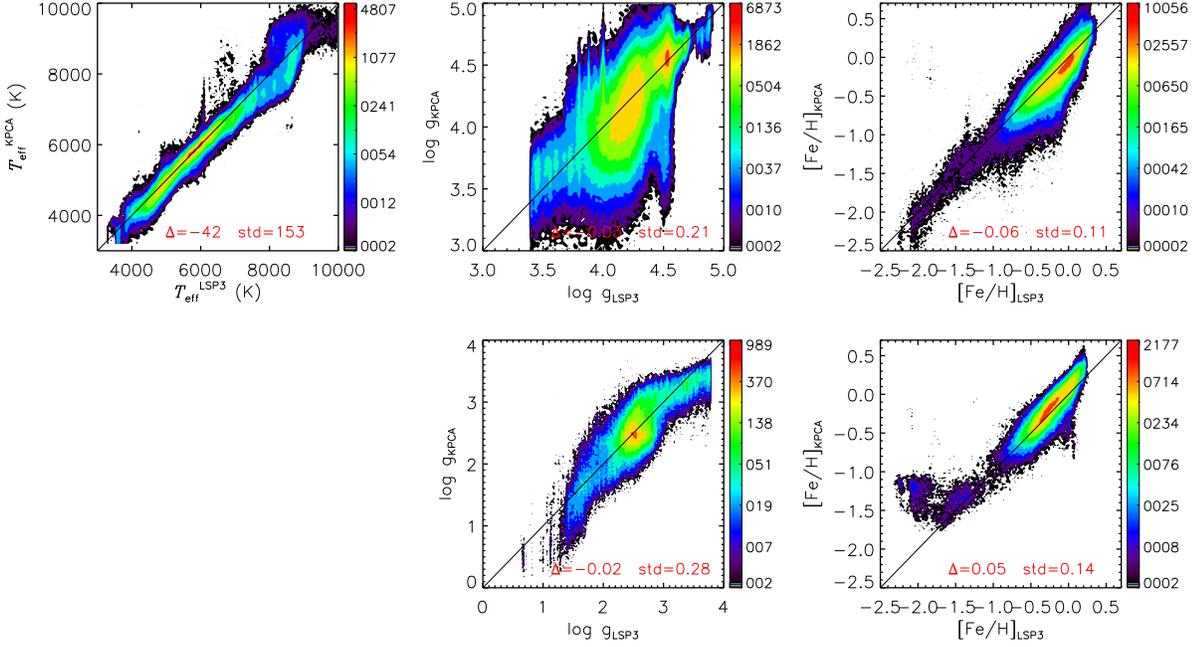}
\caption{Comparison of the estimated parameters with those derived with the weighted-mean 
method of the LSP3 for the whole LSS-GAC star sample of spectral SNR higher than 20. 
Colours indicate stellar number density in logarithmic scale.
The upper panels show results for parameters estimated with the MILES training stars, 
while the lower panels show results for log\,$g$ estimated with the LAMOST-$Kepler$ 
training set and [Fe/H] estimated with the LAMOST-APOGEE training set. 
A resistant estimate of the 
overall mean and standard deviation are marked in the plots.}
\label{Fig22}
\end{figure*}
Fig.\,22 shows the comparisons of the estimated parameters with those derived with 
the $\chi^2$-based weighted-mean method of the LSP3 for LSS-GAC DR1. 
Stars with a spectral SNR lower than 20 are used for the comparison. 
The Figure shows that on the whole, the KPCA effective 
temperatures are consistent well with the LSP3 values. The mean difference 
is only $-42$\,K, and the standard deviation for the overall stars is 153\,K. 
However, the differences show moderate trend with the temperature. 
At the higher temperature end ($T_{\rm eff}^{\rm LSP3}>7000$\,K), the KPCA 
estimates are systematically lower than the LSP3 values by about 100 -- 200\,K. 
Those lower temperature estimates are the main causes of the $-40$\,K mean difference, 
and they also contributed a significant part of the standard deviation. In fact, for 
$T_{\rm eff}^{\rm LSP3}<7000$\,K, where the systematic difference is 
negligible, value of the overall standard deviation is only 130\,K.  
Since there are few stars with temperatures higher than 7000\,K in the photometric 
sample for examination (cf. Fig.\,15), it is not straightforward to understand the 
systematic difference. It is probably that for those hot stars, the KPCA effective temperature 
are underestimated due to a yet not fully understood reason. 
Regardless of the systematic difference at $T_{\rm eff}^{\rm LSP3}>7000$\,K, 
we expected that precisions of the KPCA temperatures are comparable to those of 
the LSP3 estimates. 

The overall difference of surface gravity log\,$g$ is $-0.03\pm0.21$\,dex for dwarf stars, 
whose KPCA values are estimated with the MILES training stars,  and $-0.02\pm0.28$\,dex 
for giant stars, whose KPCA values are estimated with the LAMOST-$Kepler$ training stars. 
Though the dispersions seem not too big considering the low resolution of the 
LAMOST spectra, significant improvement has been achieved in the sense that the 
current results suffer much less systematic patterns. For the dwarf star sample, stars 
with LSP3 log\,$g$ around 4.0\,dex may have a wide range of log\,$g$ 
values (3.0 -- 4.5\,dex) in the KPCA results. 
This is because the LSP3 $\chi^2$-based weighted-mean algorithm has artificially suppressed 
log\,$g$ estimates due to the fact $\chi^2$ values are not sensitive enough 
to log\,$g$ and that log\,$g$ of the bulk dwarf/subgiant star templates covers only a 
limited range ($\sim$3.5 -- 4.5\,dex). As we increase the SNR cut of the  
sample to 100, the mean difference becomes $-0.03\pm0.19$\,dex, which changes minor  
respect to the $-0.03\pm0.21$\,dex for a SNR cut of 20, indicating that the 
differences are dominated by systematics. 
For the giant star sample, clear patterns of log\,$g$ also presented in the Figure.      
The presented patterns are consistent with those yielded by a direct comparison of the 
LSP3 weighted-mean log\,$g$ with the asteroseismic values \citep[cf. Fig.\,6 of][]{Ren+2016}. 
A similar exercise show that when we increase the SNR cut to 100, the 
mean difference becomes $-0.02\pm0.26$\,dex, again changes minor with respect to that 
of a SNR cut of 20, indicating that systematics dominate the differences.     

Both metallicities estimated using the MILES stars and the 
LAMOST-APOGEE training sets are consistent well with the LSP3 estimates except for 
very metal-poor (${\rm [Fe/H]} < -2.0$\,dex) stars or stars of super-solar metallicity.  
Note that here we have used the [M/H] for metallicity estimated using the LAMOST-APOGEE training set, 
but we adopt the tab [Fe/H] in the Figure for consistency. 
For [Fe/H] estimated using the MILES training set, the overall difference respect to the 
LSP3 values is $-0.06\pm0.11$\,dex. The $-0.06$\,dex systematic offset 
is likely caused by an overestimate of the LSP3 values \citep{Xiang+2015a}.
At the metal-rich end, the current estimates are systematically higher than the 
LSP3 values. This is a natural consequence of the underestimation 
of the LSP3 [Fe/H] due to the so-called `boundary effects' of the LSP3 
weighted-mean algorithm \citet{Xiang+2015a}, and one can see the sharp boundary in the Figure. 
At the metal-poor end, it is likely that both methods can not provide reliable [Fe/H] estimates 
due to the lack of template (training) stars of ${\rm [Fe/H]}<-2.5$\,dex. 
For [Fe/H] deduced with the LAMOST-APOGEE training set, the overall difference 
is $0.05\pm0.14$\,dex.  
The 0.05\,dex systematic offset is largely contributed by the underestimation
of the LSP3 values at the metal-rich side due to the boundary effect of the LSP3
weighted-mean algorithm. For giant stars with ${\rm [Fe/H]}_{\rm LSP3} < -2.0$\,dex, 
the KPCA estimates give a value between $-1.0$ and $-1.8$\,dex. For these stars, it is probably 
that the KPCA values are overestimated due to a lack of sufficient LAMOST-APOGEE training stars 
with ${\rm [Fe/H]}<-1.5$\,dex.   
 
\begin{figure}
\centering
\includegraphics[width=90mm]{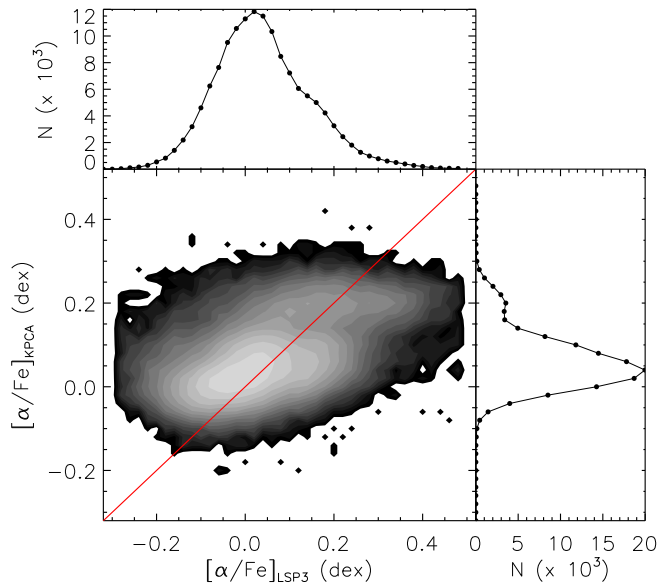}
\caption{Comparison of the estimated [$\alpha$/Fe] with those derived with the LSP3 by 
a $\chi^2$-based template matching with synthetic spectral library for LSS-GAC  
giant stars. The side panels plot the 1-dimensional stellar number density distribution of both data sets.}
\label{Fig23}
\end{figure}
We compare our [$\alpha$/Fe] estimates with those determined with a $\chi^2$-based 
template-matching method of LSP3 \citep{Liji+2016}. Fig.\,23 shows 
the comparison for LSS-GAC giant stars with a spectral SNR higher than 30. 
The Figure shows that though there is a correlation between the two sets of 
measurements, their morphologies of distribution are quite different. 
The [$\alpha$/Fe] estimated in this work have a narrower distribution, 
with only a few stars having a value smaller than $-0.1$\,dex or larger than 
0.3\,dex, while [$\alpha$/Fe] estimated with the $\chi^2$ method of LSP3 
show a significantly wider distribution. Though our current estimates are not 
externally calibrated to standard data sets, which means that systematic bias 
could be hided in our results, it is no doubt that the difference of morphology   
are largely caused by the larger random errors (0.08 -- 0.1\,dex) of [$\alpha$/Fe] 
derived with the $\chi^2$ method. We have done a test by assigning extra random 
errors of 0.05\,dex into the KPCA results, and find that the distributions are much 
more resemble to each other, but systematic trend still remains.
More examinations, especially calibration  
against high-resolution spectroscopic data sets, are necessary to further declare 
the accuracy of both the current estimates and the $\chi^2$ results.  
Nevertheless, it is remarkable that the [$\alpha$/Fe] and [$\alpha$/M] derived 
in this work have surprisingly high precisions, and the distribution of stars in the 
[M/H] -- [$\alpha$/M] as well as the [Fe/H] -- [$\alpha$/Fe] plane is quite similar to 
those from high-resolution spectroscopy \citep[e.g.][]{Venn+2004, Fuhrmann_2008, Holtzman+2015, Kordopatis+2015}.  

\section{discussion}
As has mentioned, the current work is carried out with several  
main considerations: one is to extract `weak' features that sensitive 
to parameters such as log\,$g$, [M/H], [$\alpha$/M] and individual elemental abundances 
automatically from the LAMOST spectra with the KPCA algorithm, thus to have a precise 
determination of those parameters for the millions stars of the LAMOST survey. 
Another consideration is to use a regression algorithm for parameter estimation 
to avoid potential artifacts (systematic patterns) that caused by inadequacy of 
the weighted-mean algorithm of the LSP3. 
Since the coverage of the empirical spectral templates in the parameter 
space is limited, a weighted-mean algorithm inevitably causes systematics to 
parameters of stars located near or outside the boundary of the parameter 
space. We call such an effect  the `boundary effect'  in \citet{Xiang+2015a}.  
In fact, because the empirical templates are not homogeneously distributed in the parameter space, 
stars located in the inner parameter space could also suffer systematics with the 
weighted-mean method if the weights toward individual templates are not properly assigned, 
which may cause artificial clustering of the resultant parameters. 
In addition, the current work has also, for the first time, determined absolute magnitudes 
and elemental abundances [C/H] and [N/H] from the LAMOST spectra.   

The above detailed examinations have shown that the current method has  
achieved high precisions for the estimates of $T_{\rm eff}$, 
log\,$g$, ${\rm M}_V$ and ${\rm M}_{K{\rm s}}$, [Fe/H] 
([M/H]), [$\alpha$/Fe] ([$\alpha$/M]) as well as [C/H] and [N/H] for LAMOST stars. 
Those high precisions validate the good capabilities of the KPCA algorithm for 
deducing stellar parameters from the LAMOST spectra. 
Systematic patterns hided in results of the LSP3 weighted-mean algorithm, for instance, 
the boundary effects of [Fe/H] in the metal-rich end, and the artificial suppression of log\,$g$ 
for the dwarf/subgiant stars, have been largely reduced. 
Those improvements are expected to significantly promote our capability for classification 
of stellar population, estimation of stellar distance and age, as well as the address 
of many other scientific issues. However, we emphasize that because different
training sets are used to estimate parameters for stars in different parameter ranges,
some caution is needed when interpreting the results if mixed samples
are used. More comments about this will be presented in the data release paper (Xiang et al. 2016, in preparation).

Nevertheless, there are still large space for further improvement of the parameter 
estimation. Sufficient stars with accurate, independent measurements of the stellar 
parameters that covering enough volume in the parameter space are still needed 
for either the calibration of the derived parameters or the supplementation 
of our training sets. 
For log\,$g$, the asteroseismic sample of dwarf/subgiant stars are limited to 
only a small range of both $T_{\rm eff}$ ($\sim$5400 -- 6500\,K) and [Fe/H] (dominated 
by stars of ${\rm [Fe/H]}>-1.0$\,dex). 
The current log\,$g$ estimates achieve a precision of 0.1\,dex only for 
late type ($T_{\rm eff} < 6500$\,K) and metal-rich (${\rm [Fe/H]} > -1.0$\,dex) stars, 
while for stars with higher temperature or lower metallicity, uncertainties in 
log\,$g$ estimates are significantly larger. 
For [Fe/H], both the MILES training stars and the high-resolution spectroscopy data sets 
adopted for the examination are limited to ${\rm [Fe/H]}>-2.5$\,dex, and 
the number of stars with ${\rm [Fe/H]}<-1.0$\,dex is small. What's more, the 
high-resolution spectroscopy data sets are collected from various literatures, 
and are not uniformly calibrated. 
The LAMOST-Hipparcos test stars are also limited to metal-rich stars only, thus the 
estimated absolute magnitudes for metal-poor stars need to be further examined. 
Effects from binaries, double/multiple stars as well as variable stars on the determination 
of stellar atmospheric parameters and absolute magnitudes need to be further investigated systematically. 
The number of very metal-poor stars in the LAMOST-APOGEE training set 
is still quite few. As a result, our estimates of elemental abundances for stars of 
${\rm [M/H]}<-1.5$\,dex are systematically overestimated. 
Similarly, because there are only very few stars with high [$\alpha$/Fe] ($> 0.3$\,dex),  
in our training set, [$\alpha$/Fe] for such stars are probably estimated incorrectly.     
In addition, the APOGEE abundances are determined with $\chi^2$-based     
algorithms, they may suffer from systematic biases, which have propagated 
into our deduced elemental abundances via the LAMOST-APOGEE training set. 
Systematic differences have been found between the current results and [$\alpha$/Fe] yielded by 
the template-matching method of LSP3, while further studies are needed to understand those differences. 
Further calibration with independent measurements of elemental abundances from 
high-resolution spectroscopy seems to be necessary to reduce systematic errors in the results.  

In future, parallax from {\em Gaia} \citep{Perryman+2001}, combined with metallicity and elemental 
abundances from high-resolution spectroscopy survey, e.g. the Gaia-ESO survey \citep{Gilmore+2012}, 
should be very helpful to build better training/calibration data sets for our parameter estimation from 
LAMOST spectra. More sophisticated algorithm and strategy to extract spectral features are also 
necessary to improve the measurements of elemental abundances as well as stellar atmospheric parameters.   
 
\section{Conclusion}

We have estimated stellar atmospheric parameters ($T_{\rm eff}$, log\,$g$, [Fe/H]), 
absolute magnitudes (${\rm M}_V$, ${\rm M}_{K{\rm s}}$) and elemental 
abundances ([C/H], [N/H], [$\alpha$/M], [$\alpha$/Fe]) from LAMOST spectra 
with a multivariate regression method based on kernel-based principal component analysis. 
Both internal and external examinations indicate that our method can yield 
stellar parameters from LAMOST spectra with a precision of $\sim$100\,K 
for $T_{\rm eff}$, $\sim$0.1\,dex for log\,$g$, 0.3 -- 0.4\,mag for ${\rm M}_V$ and ${\rm M}_{K{\rm s}}$, 
 $\sim$0.1\,dex for [M/H], [Fe/H], [C/H] and [N/H], and better than 0.05\,dex for [$\alpha$/M] 
([$\alpha$/Fe]), at a spectral SNR per pixel higher than 50. Even at a spectral SNR of 20, 
the precisions are still quite satisfactory, with uncertainties of $\sim$150\,K in $T_{\rm eff}$, 
0.2\,dex in log\,$g$, $\sim0.5$ -- 0.6 \,mag in ${\rm M}_V$ and ${\rm M}_{K{\rm s}}$, $\sim$0.15\,dex 
in [M/H], [Fe/H], [C/H] and [N/H],  and $\sim$0.06\,dex in [$\alpha$/M] ([$\alpha$/Fe]). 
However, we note that those numbers are generally for relatively metal-rich stars 
(${\rm [Fe/H]}>-1.5$\,dex), while parameters for more metal-poor stars are much less precise/accurate, 
and need to be used with cautious.   
With this method, systematic patterns hided in the $\chi^2$-based weighted-mean stellar atmospheric 
parameters yielded by the LSP3 can be largely reduced. 

Parameter errors for individual stars are detailedly assigned considering both  
errors induced by imperfections of the spectra and errors of the method, 
and the assigned errors are functions of the spectral SNR and stellar atmospheric parameters. 
Nevertheless, systematic offsets and biases may still exist in the results, especially in elemental abundances, 
due to a lack of external calibration against independent data sets with accurate parameter measurements, 
and simultaneously, with wide enough parameter coverage. Systematic biases in the estimated 
elemental abundances are expected to be $\sim$0.1 -- 0.2\,dex. 

The derived stellar atmospheric parameters, absolute magnitudes and elemental abundances for 
more than 1.4 million stars from the LSS-GAC, will be publicly available in the coming second 
release of value-added catalog of LSS-GAC (Xiang et al. 2016, in preparation). 
We also plan to release parameters of all stars targeted by the 
LAMOST Galactic surveys in due course. 
  
\vspace{7mm} \noindent {\bf Acknowledgments}{
This work is supported by Joint Funds of the National Natural Science Foundation of China
(Grant No.  U1531244) and the National Key Basic Research Program of China 2014CB845700.
Guoshoujing Telescope (the Large Sky Area Multi-Object Fiber
Spectroscopic Telescope LAMOST) is a National Major Scientific
Project built by the Chinese Academy of Sciences. Funding for
the project has been provided by the National Development and
Reform Commission. LAMOST is operated and managed by the National
Astronomical Observatories, Chinese Academy of Sciences. 
M.S.X. thanks Prof. Chao Liu for helpful discussion.}

\bibliographystyle{mn2e}
\bibliography{kpca}

\begin{thebibliography}{75}
\expandafter\ifx\csname natexlab\endcsname\relax\def\natexlab#1{#1}\fi

\bibitem[{{Allende Prieto} {et~al}\mbox{.}(2006){Allende Prieto}, {Beers},
  {Wilhelm}, {Newberg}, {Rockosi}, {Yanny}, \& {Lee}}]{Allende_Prieto+2006}
{Allende Prieto} C., {Beers} T.~C., {Wilhelm} R., {Newberg} H.~J., {Rockosi}
  C.~M., {Yanny} B., {Lee} Y.~S., 2006, \apj, 636, 804

\bibitem[{{Anderson} \& {Francis}(2012)}]{Anderson+2012}
{Anderson} E., {Francis} C., 2012, Astronomy Letters, 38, 331

\bibitem[{{Bailer-Jones} {et~al}\mbox{.}(2013){Bailer-Jones}, {Andrae},
  {Arcay}, {Astraatmadja}, {Bellas-Velidis}, {Berihuete}, {Bijaoui},
  {Carri{\'o}n}, {Dafonte}, {Damerdji}, {Dapergolas}, {de Laverny},
  {Delchambre}, {Drazinos}, {Drimmel}, {Fr{\'e}mat}, {Fustes},
  {Garc{\'{\i}}a-Torres}, {Gu{\'e}d{\'e}}, {Heiter}, {Janotto}, {Karampelas},
  {Kim}, {Knude}, {Kolka}, {Kontizas}, {Kontizas}, {Korn}, {Lanzafame},
  {Lebreton}, {Lindstr{\o}m}, {Liu}, {Livanou}, {Lobel}, {Manteiga},
  {Martayan}, {Ordenovic}, {Pichon}, {Recio-Blanco}, {Rocca-Volmerange},
  {Sarro}, {Smith}, {Sordo}, {Soubiran}, {Surdej}, {Th{\'e}venin},
  {Tsalmantza}, {Vallenari}, \& {Zorec}}]{Bailer-Jones+2013}
{Bailer-Jones} C.~A.~L. {et~al.}, 2013, \aap, 559, A74

\bibitem[{{Bensby}, {Feltzing} \& {Lundstr{\"o}m}(2003){Bensby}, {Feltzing}, \&
  {Lundstr{\"o}m}}]{Bensby+2003}
{Bensby} T., {Feltzing} S., {Lundstr{\"o}m} I., 2003, \aap, 410, 527

\bibitem[{{Bu} \& {Pan}(2015)}]{Bu+2015}
{Bu} Y., {Pan} J., 2015, \mnras, 447, 256

\bibitem[{{Cenarro} {et~al}\mbox{.}(2007){Cenarro}, {Peletier},
  {S{\'a}nchez-Bl{\'a}zquez}, {Selam}, {Toloba}, {Cardiel},
  {Falc{\'o}n-Barroso}, {Gorgas}, {Jim{\'e}nez-Vicente}, \&
  {Vazdekis}}]{Cenarro+2007}
{Cenarro} A.~J. {et~al.}, 2007, \mnras, 374, 664

\bibitem[{{Cui} {et~al}\mbox{.}(2012){Cui}, {Zhao}, {Chu}, {Li}, {Li}, {Zhang},
  {Su}, {Yao}, {Wang}, {Xing}, {Li}, {Zhu}, {Wang}, {Gu}, {Luo}, {Xu}, {Zhang},
  {Liu}, {Zhang}, {Yang}, {Cao}, {Chen}, {Chen}, {Chen}, {Chen}, {Chu}, {Feng},
  {Gong}, {Hou}, {Hu}, {Hu}, {Hu}, {Jia}, {Jiang}, {Jiang}, {Jiang}, {Jin},
  {Li}, {Li}, {Li}, {Liu}, {Liu}, {Lu}, {Mao}, {Men}, {Qi}, {Qi}, {Shi},
  {Tang}, {Tao}, {Wang}, {Wang}, {Wang}, {Wang}, {Wang}, {Wang}, {Wang},
  {Wang}, {Wang}, {Wang}, {Wang}, {Wang}, {Xu}, {Xu}, {Yang}, {Yu}, {Yuan},
  {Yuan}, {Zhai}, {Zhang}, {Zhang}, {Zhang}, {Zhao}, {Zhou}, {Zhou}, {Zhu}, \&
  {Zou}}]{Cui+2012}
{Cui} X.-Q. {et~al.}, 2012, Research in Astronomy and Astrophysics, 12, 1197

\bibitem[{{De Cat} {et~al}\mbox{.}(2015){De Cat}, {Fu}, {Ren}, {Yang}, {Shi},
  {Luo}, {Yang}, {Wang}, {Zhang}, {Shi}, {Zhang}, {Dong}, {Catanzaro},
  {Corbally}, {Frasca}, {Gray}, {Molenda-{\.Z}akowicz}, {Uytterhoeven},
  {Briquet}, {Bruntt}, {Frandsen}, {Kiss}, {Kurtz}, {Marconi}, {Niemczura},
  {{\O}stensen}, {Ripepi}, {Smalley}, {Southworth}, {Szab{\'o}}, {Telting},
  {Karoff}, {Silva Aguirre}, {Wu}, {Hou}, {Jin}, \& {Zhou}}]{De_Cat+2015}
{De Cat} P. {et~al.}, 2015, \apjs, 220, 19

\bibitem[{{Demarque} {et~al}\mbox{.}(2004){Demarque}, {Woo}, {Kim}, \&
  {Yi}}]{Demarque+2004}
{Demarque} P., {Woo} J.-H., {Kim} Y.-C., {Yi} S.~K., 2004, \apjs, 155, 667

\bibitem[{{Deng} {et~al}\mbox{.}(2012){Deng}, {Newberg}, {Liu}, {Carlin},
  {Beers}, {Chen}, {Chen}, {Christlieb}, {Grillmair}, {Guhathakurta}, {Han},
  {Hou}, {Lee}, {L{\'e}pine}, {Li}, {Liu}, {Pan}, {Sellwood}, {Wang}, {Wang},
  {Yang}, {Yanny}, {Zhang}, {Zhang}, {Zheng}, \& {Zhu}}]{Deng+2012}
{Deng} L.-C. {et~al.}, 2012, Research in Astronomy and Astrophysics, 12, 735

\bibitem[{{Dotter} {et~al}\mbox{.}(2008){Dotter}, {Chaboyer}, {Jevremovi{\'c}},
  {Kostov}, {Baron}, \& {Ferguson}}]{Dotter+2008}
{Dotter} A., {Chaboyer} B., {Jevremovi{\'c}} D., {Kostov} V., {Baron} E.,
  {Ferguson} J.~W., 2008, \apjs, 178, 89

\bibitem[{{Falc{\'o}n-Barroso} {et~al}\mbox{.}(2011){Falc{\'o}n-Barroso},
  {S{\'a}nchez-Bl{\'a}zquez}, {Vazdekis}, {Ricciardelli}, {Cardiel}, {Cenarro},
  {Gorgas}, \& {Peletier}}]{Falcon-Barroso+2011}
{Falc{\'o}n-Barroso} J., {S{\'a}nchez-Bl{\'a}zquez} P., {Vazdekis} A.,
  {Ricciardelli} E., {Cardiel} N., {Cenarro} A.~J., {Gorgas} J., {Peletier}
  R.~F., 2011, \aap, 532, A95

\bibitem[{{Freeman}(2012)}]{Freeman2012}
{Freeman} K.~C., 2012, in Astronomical Society of the Pacific Conference
  Series, Vol. 458, Galactic Archaeology: Near-Field Cosmology and the
  Formation of the Milky Way, {Aoki} W., {Ishigaki} M., {Suda} T., {Tsujimoto}
  T., {Arimoto} N., eds., p. 393

\bibitem[{{Fuhrmann}(2008)}]{Fuhrmann_2008}
{Fuhrmann} K., 2008, \mnras, 384, 173

\bibitem[{{Gao} {et~al}\mbox{.}(2015){Gao}, {Zhang}, {Xiang}, {Huang}, {Liu},
  {Luo}, {Zhang}, {Wu}, {Zhang}, {Li}, \& {Du}}]{Gao+2015}
{Gao} H. {et~al.}, 2015, Research in Astronomy and Astrophysics, 15, 2204

\bibitem[{{Garc{\'{\i}}a P{\'e}rez} {et~al}\mbox{.}(2016){Garc{\'{\i}}a
  P{\'e}rez}, {Allende Prieto}, {Holtzman}, {Shetrone}, {M{\'e}sz{\'a}ros},
  {Bizyaev}, {Carrera}, {Cunha}, {Garc{\'{\i}}a-Hern{\'a}ndez}, {Johnson},
  {Majewski}, {Nidever}, {Schiavon}, {Shane}, {Smith}, {Sobeck}, {Troup},
  {Zamora}, {Weinberg}, {Bovy}, {Eisenstein}, {Feuillet}, {Frinchaboy},
  {Hayden}, {Hearty}, {Nguyen}, {OConnell}, {Pinsonneault}, {Wilson}, \&
  {Zasowski}}]{Garcia_Perez+2015}
{Garc{\'{\i}}a P{\'e}rez} A.~E. {et~al.}, 2016, \aj, 151, 144

\bibitem[{{Gilmore} {et~al}\mbox{.}(2012){Gilmore}, {Randich}, {Asplund},
  {Binney}, {Bonifacio}, {Drew}, {Feltzing}, {Ferguson}, {Jeffries}, {Micela},
  \& et~al.}]{Gilmore+2012}
{Gilmore} G. {et~al.}, 2012, The Messenger, 147, 25

\bibitem[{{Hekker} {et~al}\mbox{.}(2013){Hekker}, {Elsworth}, {Mosser},
  {Kallinger}, {Basu}, {Chaplin}, \& {Stello}}]{Hekker+2013}
{Hekker} S., {Elsworth} Y., {Mosser} B., {Kallinger} T., {Basu} S., {Chaplin}
  W.~J., {Stello} D., 2013, \aap, 556, A59

\bibitem[{{Henden} \& {Munari}(2014)}]{Henden+2014}
{Henden} A., {Munari} U., 2014, Contributions of the Astronomical Observatory
  Skalnate Pleso, 43, 518

\bibitem[{{Holtzman} {et~al}\mbox{.}(2015){Holtzman}, {Shetrone}, {Johnson},
  {Allende Prieto}, {Anders}, {Andrews}, {Beers}, {Bizyaev}, {Blanton}, {Bovy},
  {Carrera}, {Chojnowski}, {Cunha}, {Eisenstein}, {Feuillet}, {Frinchaboy},
  {Galbraith-Frew}, {Garc{\'{\i}}a P{\'e}rez}, {Garc{\'{\i}}a-Hern{\'a}ndez},
  {Hasselquist}, {Hayden}, {Hearty}, {Ivans}, {Majewski}, {Martell},
  {Meszaros}, {Muna}, {Nidever}, {Nguyen}, {O Connell}, {Pan}, {Pinsonneault},
  {Robin}, {Schiavon}, {Shane}, {Sobeck}, {Smith}, {Troup}, {Weinberg},
  {Wilson}, {Wood-Vasey}, {Zamora}, \& {Zasowski}}]{Holtzman+2015}
{Holtzman} J.~A. {et~al.}, 2015, \aj, 150, 148

\bibitem[{{Hou} {et~al}\mbox{.}(2013){Hou}, {Zhong}, {Chen}, {Yu}, {Liu}, \&
  {Deng}}]{Hou+2013}
{Hou} J.~L., {Zhong} J., {Chen} L., {Yu} J.~C., {Liu} C., {Deng} L.~C., 2013,
  in IAU Symposium, Vol. 292, IAU Symposium, {Wong} T., {Ott} J., eds., pp.
  105--105

\bibitem[{{Huang} {et~al}\mbox{.}(2015{\natexlab{a}}){Huang}, {Liu}, {Yuan},
  {Xiang}, {Chen}, \& {Zhang}}]{Huang+2015b}
{Huang} Y., {Liu} X.-W., {Yuan} H.-B., {Xiang} M.-S., {Chen} B.-Q., {Zhang}
  H.-W., 2015{\natexlab{a}}, \mnras, 454, 2863

\bibitem[{{Huang} {et~al}\mbox{.}(2015{\natexlab{b}}){Huang}, {Liu}, {Zhang},
  {Yuan}, {Xiang}, {Chen}, {Ren}, {Sun}, {Wang}, {Zhang}, {Hou}, {Wang}, \&
  {Yang}}]{Huang+2015a}
{Huang} Y. {et~al.}, 2015{\natexlab{b}}, Research in Astronomy and
  Astrophysics, 15, 1240

\bibitem[{{Huber} {et~al}\mbox{.}(2014){Huber}, {Silva Aguirre}, {Matthews},
  {Pinsonneault}, {Gaidos}, {Garc{\'{\i}}a}, {Hekker}, {Mathur}, {Mosser},
  {Torres}, {Bastien}, {Basu}, {Bedding}, {Chaplin}, {Demory}, {Fleming},
  {Guo}, {Mann}, {Rowe}, {Serenelli}, {Smith}, \& {Stello}}]{Huber+2014}
{Huber} D. {et~al.}, 2014, \apjs, 211, 2

\bibitem[{{Jacobson}, {Pilachowski} \& {Friel}(2011){Jacobson}, {Pilachowski},
  \& {Friel}}]{Jacobson+2011}
{Jacobson} H.~R., {Pilachowski} C.~A., {Friel} E.~D., 2011, \aj, 142, 59

\bibitem[{{Jofr{\'e}} {et~al}\mbox{.}(2015){Jofr{\'e}}, {Heiter}, {Soubiran},
  {Blanco-Cuaresma}, {Masseron}, {Nordlander}, {Chemin}, {Worley}, {Van Eck},
  {Hourihane}, {Gilmore}, {Adibekyan}, {Bergemann}, {Cantat-Gaudin},
  {Delgado-Mena}, {Gonz{\'a}lez Hern{\'a}ndez}, {Guiglion}, {Lardo}, {de
  Laverny}, {Lind}, {Magrini}, {Mikolaitis}, {Montes}, {Pancino},
  {Recio-Blanco}, {Sordo}, {Sousa}, {Tabernero}, \& {Vallenari}}]{Jofre+2015}
{Jofr{\'e}} P. {et~al.}, 2015, \aap, 582, A81

\bibitem[{{Koleva} {et~al}\mbox{.}(2009){Koleva}, {Prugniel}, {Bouchard}, \&
  {Wu}}]{Koleva+2009}
{Koleva} M., {Prugniel} P., {Bouchard} A., {Wu} Y., 2009, \aap, 501, 1269

\bibitem[{{Kordopatis} {et~al}\mbox{.}(2013){Kordopatis}, {Gilmore},
  {Steinmetz}, {Boeche}, {Seabroke}, {Siebert}, {Zwitter}, {Binney}, {de
  Laverny}, {Recio-Blanco}, {Williams}, {Piffl}, {Enke}, {Roeser}, {Bijaoui},
  {Wyse}, {Freeman}, {Munari}, {Carrillo}, {Anguiano}, {Burton}, {Campbell},
  {Cass}, {Fiegert}, {Hartley}, {Parker}, {Reid}, {Ritter}, {Russell},
  {Stupar}, {Watson}, {Bienaym{\'e}}, {Bland-Hawthorn}, {Gerhard}, {Gibson},
  {Grebel}, {Helmi}, {Navarro}, {Conrad}, {Famaey}, {Faure}, {Just}, {Kos},
  {Matijevi{\v c}}, {McMillan}, {Minchev}, {Scholz}, {Sharma}, {Siviero}, {de
  Boer}, \& {{\v Z}erjal}}]{Kordopatis+2013}
{Kordopatis} G. {et~al.}, 2013, \aj, 146, 134

\bibitem[{{Kordopatis} {et~al}\mbox{.}(2015){Kordopatis}, {Wyse}, {Gilmore},
  {Recio-Blanco}, {de Laverny}, {Hill}, {Adibekyan}, {Heiter}, {Minchev},
  {Famaey}, {Bensby}, {Feltzing}, {Guiglion}, {Korn}, {Mikolaitis},
  {Schultheis}, {Vallenari}, {Bayo}, {Carraro}, {Flaccomio}, {Franciosini},
  {Hourihane}, {Jofr{\'e}}, {Koposov}, {Lardo}, {Lewis}, {Lind}, {Magrini},
  {Morbidelli}, {Pancino}, {Randich}, {Sacco}, {Worley}, \&
  {Zaggia}}]{Kordopatis+2015}
{Kordopatis} G. {et~al.}, 2015, \aap, 582, A122

\bibitem[{{Lee} {et~al}\mbox{.}(2008{\natexlab{a}}){Lee}, {Beers}, {Sivarani},
  {Allende Prieto}, {Koesterke}, {Wilhelm}, {Re Fiorentin}, {Bailer-Jones},
  {Norris}, {Rockosi}, {Yanny}, {Newberg}, {Covey}, {Zhang}, \&
  {Luo}}]{Lee+2008a}
{Lee} Y.~S. {et~al.}, 2008{\natexlab{a}}, \aj, 136, 2022

\bibitem[{{Lee} {et~al}\mbox{.}(2008{\natexlab{b}}){Lee}, {Beers}, {Sivarani},
  {Johnson}, {An}, {Wilhelm}, {Allende Prieto}, {Koesterke}, {Re Fiorentin},
  {Bailer-Jones}, {Norris}, {Yanny}, {Rockosi}, {Newberg}, {Cudworth}, \&
  {Pan}}]{Lee+2008b}
{Lee} Y.~S. {et~al.}, 2008{\natexlab{b}}, \aj, 136, 2050

\bibitem[{{Li} {et~al}\mbox{.}(2016){Li}, {Han}, {Xiang}, {Shi}, {Zhao}, {Liu},
  {Zhang}, {Yuan}, {Ci}, {Zhang}, {Wang}, {Huang}, {Zhang}, {Hou}, {Wang}, \&
  {Cao}}]{Liji+2016}
{Li} J. {et~al.}, 2016, Research in Astronomy and Astrophysics, 16, 010

\bibitem[{{Li} {et~al}\mbox{.}(2015){Li}, {Lu}, {Comte}, {Luo}, {Zhao}, \&
  {Wang}}]{Li_Lu+2015}
{Li} X., {Lu} Y., {Comte} G., {Luo} A., {Zhao} Y., {Wang} Y., 2015, \apjs, 218,
  3

\bibitem[{{Liu} {et~al}\mbox{.}(2015){Liu}, {Fang}, {Wu}, {Deng}, {Wang},
  {Wang}, {Fu}, {Hou}, {Li}, \& {Zhang}}]{Liuchao+2015}
{Liu} C. {et~al.}, 2015, \apj, 807, 4

\bibitem[{{Liu} {et~al}\mbox{.}(2014){Liu}, {Yuan}, {Huo}, {Deng}, {Hou},
  {Zhao}, {Zhao}, {Shi}, {Luo}, {Xiang}, {Zhang}, {Huang}, \&
  {Zhang}}]{Liu+2014}
{Liu} X.-W. {et~al.}, 2014, in IAU Symposium, Vol. 298, IAU Symposium,
  {Feltzing} S., {Zhao} G., {Walton} N.~A., {Whitelock} P., eds., pp. 310--321

\bibitem[{{Liu}, {Zhao} \& {Hou}(2015){Liu}, {Zhao}, \& {Hou}}]{Liu+2015}
{Liu} X.-W., {Zhao} G., {Hou} J.-L., 2015, Research in Astronomy and
  Astrophysics, 15, 1089

\bibitem[{{Lu} \& {Li}(2015)}]{Lu_Li+2015}
{Lu} Y., {Li} X., 2015, \mnras, 452, 1394

\bibitem[{{Luo} {et~al}\mbox{.}(2012){Luo}, {Zhang}, {Zhao}, {Zhao}, {Cui},
  {Li}, {Chu}, {Shi}, {Wang}, {Zhang}, {Bai}, {Chen}, {Wang}, {Guo}, {Chen},
  {Du}, {Kong}, {Lei}, {Li}, {Song}, {Wu}, {Zhang}, {Zhou}, {Zuo}, {Du}, {He},
  {Hou}, {Dong}, {Li}, {Li}, {Li}, {Song}, {Tian}, {Wang}, {Wu}, {Yang},
  {Yuan}, {Cao}, {Chen}, {Chen}, {Chen}, {Chu}, {Feng}, {Gong}, {Gu}, {Hou},
  {Huo}, {Hu}, {Hu}, {Hu}, {Jia}, {Jiang}, {Jiang}, {Jiang}, {Jin}, {Li}, {Li},
  {Li}, {Li}, {Li}, {Liu}, {Liu}, {Liu}, {Lu}, {Lu}, {Luo}, {Mao}, {Men}, {Ni},
  {Qi}, {Qi}, {Shi}, {Su}, {Sun}, {Su}, {Tang}, {Tao}, {Tu}, {Wang}, {Wang},
  {Wang}, {Wang}, {Wang}, {Wang}, {Wang}, {Wang}, {Wang}, {Wang}, {Wang},
  {Wang}, {Wang}, {Wang}, {Wei}, {Xue}, {Xing}, {Xu}, {Xu}, {Xu}, {Yang},
  {Yang}, {Yao}, {Yu}, {Yuan}, {Zhai}, {Zhang}, {Zhang}, {Zhang}, {Zhang},
  {Zhang}, {Zhang}, {Zhao}, {Zhou}, {Zhu}, {Zhu}, \& {Zou}}]{Luo+2012}
{Luo} A.-L. {et~al.}, 2012, Research in Astronomy and Astrophysics, 12, 1243

\bibitem[{{Luo} {et~al}\mbox{.}(2015){Luo}, {Zhao}, {Zhao}, {Deng}, {Liu},
  {Jing}, {Wang}, {Zhang}, {Shi}, {Cui}, {Chu}, {Li}, {Bai}, {Wu}, {Cai},
  {Cao}, {Cao}, {Carlin}, {Chen}, {Chen}, {Chen}, {Chen}, {Chen}, {Chen},
  {Chen}, {Christlieb}, {Chu}, {Cui}, {Dong}, {Du}, {Fan}, {Feng}, {Fu}, {Gao},
  {Gong}, {Gu}, {Guo}, {Han}, {He}, {Hou}, {Hou}, {Hou}, {Hu}, {Hu}, {Hu},
  {Huo}, {Jia}, {Jiang}, {Jiang}, {Jiang}, {Jin}, {Kong}, {Kong}, {Lei}, {Li},
  {Li}, {Li}, {Li}, {Li}, {Li}, {Li}, {Li}, {Li}, {Li}, {Li}, {Li}, {Liang},
  {Lin}, {Liu}, {Liu}, {Liu}, {Liu}, {Lu}, {Luo}, {Mao}, {Newberg}, {Ni}, {Qi},
  {Qi}, {Shen}, {Shi}, {Song}, {Song}, {Su}, {Su}, {Tang}, {Tao}, {Tian},
  {Wang}, {Wang}, {Wang}, {Wang}, {Wang}, {Wang}, {Wang}, {Wang}, {Wang},
  {Wang}, {Wang}, {Wang}, {Wang}, {Wang}, {Wang}, {Wang}, {Wang}, {Wang},
  {Wang}, {Wang}, {Wei}, {Wei}, {Wu}, {Wu}, {Wu}, {Wu}, {Xing}, {Xu}, {Xu},
  {Xu}, {Yan}, {Yang}, {Yang}, {Yang}, {Yang}, {Yao}, {Yu}, {Yuan}, {Yuan},
  {Yuan}, {Yuan}, {Zhai}, {Zhang}, {Zhang}, {Zhang}, {Zhang}, {Zhang}, {Zhang},
  {Zhang}, {Zhang}, {Zhao}, {Zhou}, {Zhou}, {Zhu}, {Zhu}, {Zou}, \&
  {Zuo}}]{Luo+2015}
{Luo} A.-L. {et~al.}, 2015, Research in Astronomy and Astrophysics, 15, 1095

\bibitem[{{Majewski} {et~al}\mbox{.}(2010){Majewski}, {Wilson}, {Hearty},
  {Schiavon}, \& {Skrutskie}}]{Majewski+2010}
{Majewski} S.~R., {Wilson} J.~C., {Hearty} F., {Schiavon} R.~R., {Skrutskie}
  M.~F., 2010, in IAU Symposium, Vol. 265, IAU Symposium, {Cunha} K., {Spite}
  M., {Barbuy} B., eds., pp. 480--481

\bibitem[{{Manteiga} {et~al}\mbox{.}(2010){Manteiga}, {Ord{\'o}{\~n}ez},
  {Dafonte}, \& {Arcay}}]{Manteiga+2010}
{Manteiga} M., {Ord{\'o}{\~n}ez} D., {Dafonte} C., {Arcay} B., 2010, \pasp,
  122, 608

\bibitem[{{M\"uller} {et~al}\mbox{.}(2001){M\"uller}, {Mika}, {R\"atsch},
  {Tsuda}, \& {Sch\"olkopf}}]{Muller+2001}
{M\"uller} K.~R., {Mika} S., {R\"atsch} G., {Tsuda} K., {Sch\"olkopf} B., 2001,
  IEEE Transactions on Neural Networks, 12, 181

\bibitem[{{Ness} {et~al}\mbox{.}(2015){Ness}, {Hogg}, {Rix}, {Ho}, \&
  {Zasowski}}]{Ness+2015}
{Ness} M., {Hogg} D.~W., {Rix} H.-W., {Ho} A.~Y.~Q., {Zasowski} G., 2015, \apj,
  808, 16

\bibitem[{{Perryman} {et~al}\mbox{.}(2001){Perryman}, {de Boer}, {Gilmore},
  {H{\o}g}, {Lattanzi}, {Lindegren}, {Luri}, {Mignard}, {Pace}, \& {de
  Zeeuw}}]{Perryman+2001}
{Perryman} M.~A.~C. {et~al.}, 2001, \aap, 369, 339

\bibitem[{{Perryman} {et~al}\mbox{.}(1997){Perryman}, {Lindegren},
  {Kovalevsky}, {Hoeg}, {Bastian}, {Bernacca}, {Cr{\'e}z{\'e}}, {Donati},
  {Grenon}, {Grewing}, {van Leeuwen}, {van der Marel}, {Mignard}, {Murray}, {Le
  Poole}, {Schrijver}, {Turon}, {Arenou}, {Froeschl{\'e}}, \&
  {Petersen}}]{Perryman+1997}
{Perryman} M.~A.~C. {et~al.}, 1997, \aap, 323, L49

\bibitem[{{Prugniel} {et~al}\mbox{.}(2007){Prugniel}, {Soubiran}, {Koleva}, \&
  {Le Borgne}}]{Prugniel+2007}
{Prugniel} P., {Soubiran} C., {Koleva} M., {Le Borgne} D., 2007, ArXiv
  Astrophysics e-prints, 0703658

\bibitem[{{Re Fiorentin} {et~al}\mbox{.}(2007){Re Fiorentin}, {Bailer-Jones},
  {Lee}, {Beers}, {Sivarani}, {Wilhelm}, {Allende Prieto}, \&
  {Norris}}]{Re_Fiorentin+2007}
{Re Fiorentin} P., {Bailer-Jones} C.~A.~L., {Lee} Y.~S., {Beers} T.~C.,
  {Sivarani} T., {Wilhelm} R., {Allende Prieto} C., {Norris} J.~E., 2007, \aap,
  467, 1373

\bibitem[{{Recio-Blanco}, {Bijaoui} \& {de Laverny}(2006){Recio-Blanco},
  {Bijaoui}, \& {de Laverny}}]{Recio-Blanco+2006}
{Recio-Blanco} A., {Bijaoui} A., {de Laverny} P., 2006, \mnras, 370, 141

\bibitem[{{Recio-Blanco} {et~al}\mbox{.}(2016){Recio-Blanco}, {de Laverny},
  {Allende Prieto}, {Fustes}, {Manteiga}, {Arcay}, {Bijaoui}, {Dafonte},
  {Ordenovic}, \& {Ordo{\~n}ez Blanco}}]{Recio-Blanco+2015}
{Recio-Blanco} A. {et~al.}, 2016, \aap, 585, A93

\bibitem[{{Ren} {et~al}\mbox{.}(2016){Ren}, {Liu}, {Xiang}, {Huang}, {Hekker},
  {Wang}, {Yuan}, {Rebassa-Mansergas}, {Chen}, {Sun}, {Zhang}, {Huo}, {Zhang},
  {Zhang}, {Hou}, \& {Wang}}]{Ren+2016}
{Ren} J.-J. {et~al.}, 2016, Research in Astronomy and Astrophysics, 16, 009

\bibitem[{{Richer} {et~al}\mbox{.}(1998){Richer}, {Fahlman}, {Rosvick}, \&
  {Ibata}}]{Richer+1998}
{Richer} H.~B., {Fahlman} G.~G., {Rosvick} J., {Ibata} R., 1998, \apjl, 504,
  L91

\bibitem[{{Salaris}, {Weiss} \& {Percival}(2004){Salaris}, {Weiss}, \&
  {Percival}}]{Salaris+2004}
{Salaris} M., {Weiss} A., {Percival} S.~M., 2004, \aap, 414, 163

\bibitem[{{S{\'a}nchez-Bl{\'a}zquez}
  {et~al}\mbox{.}(2006){S{\'a}nchez-Bl{\'a}zquez}, {Peletier},
  {Jim{\'e}nez-Vicente}, {Cardiel}, {Cenarro}, {Falc{\'o}n-Barroso}, {Gorgas},
  {Selam}, \& {Vazdekis}}]{Sanchez-Blazquez+2006}
{S{\'a}nchez-Bl{\'a}zquez} P. {et~al.}, 2006, \mnras, 371, 703

\bibitem[{{Schlegel}, {Finkbeiner} \& {Davis}(1998){Schlegel}, {Finkbeiner}, \&
  {Davis}}]{Schlegel+1998}
{Schlegel} D.~J., {Finkbeiner} D.~P., {Davis} M., 1998, \apj, 500, 525

\bibitem[{{Sch{\"o}lpokf}, {Smola} \& {M{\"u}ller}(1998){Sch{\"o}lpokf},
  {Smola}, \& {M{\"u}ller}}]{Scholkopf+1998}
{Sch{\"o}lpokf} B., {Smola} A.~J., {M{\"u}ller} K.~R., 1998, Neural
  Computation, 10, 1299

\bibitem[{{Skrutskie} {et~al}\mbox{.}(2006){Skrutskie}, {Cutri}, {Stiening},
  {Weinberg}, {Schneider}, {Carpenter}, {Beichman}, {Capps}, {Chester},
  {Elias}, {Huchra}, {Liebert}, {Lonsdale}, {Monet}, {Price}, {Seitzer},
  {Jarrett}, {Kirkpatrick}, {Gizis}, {Howard}, {Evans}, {Fowler}, {Fullmer},
  {Hurt}, {Light}, {Kopan}, {Marsh}, {McCallon}, {Tam}, {Van Dyk}, \&
  {Wheelock}}]{Skrutskie+2006}
{Skrutskie} M.~F. {et~al.}, 2006, \aj, 131, 1163

\bibitem[{{Soubiran} {et~al}\mbox{.}(2010){Soubiran}, {Le Campion}, {Cayrel de
  Strobel}, \& {Caillo}}]{Soubiran+2010}
{Soubiran} C., {Le Campion} J.-F., {Cayrel de Strobel} G., {Caillo} A., 2010,
  \aap, 515, A111

\bibitem[{{Steinmetz} {et~al}\mbox{.}(2006){Steinmetz}, {Zwitter}, {Siebert},
  {Watson}, {Freeman}, {Munari}, {Campbell}, {Williams}, {Seabroke}, {Wyse},
  {Parker}, {Bienaym{\'e}}, {Roeser}, {Gibson}, {Gilmore}, {Grebel}, {Helmi},
  {Navarro}, {Burton}, {Cass}, {Dawe}, {Fiegert}, {Hartley}, {Russell},
  {Saunders}, {Enke}, {Bailin}, {Binney}, {Bland-Hawthorn}, {Boeche}, {Dehnen},
  {Eisenstein}, {Evans}, {Fiorucci}, {Fulbright}, {Gerhard}, {Jauregi}, {Kelz},
  {Mijovi{\'c}}, {Minchev}, {Parmentier}, {Pe{\~n}arrubia}, {Quillen}, {Read},
  {Ruchti}, {Scholz}, {Siviero}, {Smith}, {Sordo}, {Veltz}, {Vidrih}, {von
  Berlepsch}, {Boyle}, \& {Schilbach}}]{Steinmetz+2006}
{Steinmetz} M. {et~al.}, 2006, \aj, 132, 1645

\bibitem[{{Venn} {et~al}\mbox{.}(2004){Venn}, {Irwin}, {Shetrone}, {Tout},
  {Hill}, \& {Tolstoy}}]{Venn+2004}
{Venn} K.~A., {Irwin} M., {Shetrone} M.~D., {Tout} C.~A., {Hill} V., {Tolstoy}
  E., 2004, \aj, 128, 1177

\bibitem[{{Wang} {et~al}\mbox{.}(1996){Wang}, {Su}, {Chu}, {Cui}, \&
  {Wang}}]{Wang+1996}
{Wang} S.-G., {Su} D.-Q., {Chu} Y.-Q., {Cui} X., {Wang} Y.-N., 1996, \ao, 35,
  5155

\bibitem[{{Wenger} {et~al}\mbox{.}(2000){Wenger}, {Ochsenbein}, {Egret},
  {Dubois}, {Bonnarel}, {Borde}, {Genova}, {Jasniewicz}, {Lalo{\"e}},
  {Lesteven}, \& {Monier}}]{Wenger+2000}
{Wenger} M. {et~al.}, 2000, \aaps, 143, 9

\bibitem[{{Wu} {et~al}\mbox{.}(2014){Wu}, {Du}, {Luo}, {Zhao}, \&
  {Yuan}}]{Wu+2014}
{Wu} Y., {Du} B., {Luo} A., {Zhao} Y., {Yuan} H., 2014, in IAU Symposium, Vol.
  306, Statistical Challenges in 21st Century Cosmology, {Heavens} A., {Starck}
  J.-L., {Krone-Martins} A., eds., pp. 340--342

\bibitem[{{Wu} {et~al}\mbox{.}(2011){Wu}, {Luo}, {Li}, {Shi}, {Prugniel},
  {Liang}, {Zhao}, {Zhang}, {Bai}, {Wei}, {Dong}, {Zhang}, \& {Chen}}]{Wu+2011}
{Wu} Y. {et~al.}, 2011, Research in Astronomy and Astrophysics, 11, 924

\bibitem[{{Xiang} {et~al}\mbox{.}(2015{\natexlab{a}}){Xiang}, {Liu}, {Yuan},
  {Huang}, {Huo}, {Zhang}, {Chen}, {Zhang}, {Sun}, {Wang}, {Zhao}, {Shi},
  {Luo}, {Li}, {Wu}, {Bai}, {Zhang}, {Hou}, {Yuan}, {Li}, \&
  {Wei}}]{Xiang+2015a}
{Xiang} M.~S. {et~al.}, 2015{\natexlab{a}}, \mnras, 448, 822

\bibitem[{{Xiang} {et~al}\mbox{.}(2015{\natexlab{b}}){Xiang}, {Liu}, {Yuan},
  {Huang}, {Wang}, {Ren}, {Chen}, {Sun}, {Zhang}, {Huo}, \&
  {Rebassa-Mansergas}}]{Xiang+2015c}
{Xiang} M.-S. {et~al.}, 2015{\natexlab{b}}, Research in Astronomy and
  Astrophysics, 15, 1209

\bibitem[{{Xiang} {et~al}\mbox{.}(2015{\natexlab{c}}){Xiang}, {Liu}, {Yuan},
  {Huo}, {Huang}, {Zheng}, {Zhang}, {Chen}, {Zhang}, {Sun}, {Wang}, {Zhao},
  {Shi}, {Luo}, {Li}, {Bai}, {Zhang}, {Hou}, {Yuan}, \& {Li}}]{Xiang+2015b}
{Xiang} M.~S. {et~al.}, 2015{\natexlab{c}}, \mnras, 448, 90

\bibitem[{{Yang} \& {Li}(2015)}]{Yang_Li+2015}
{Yang} T., {Li} X., 2015, \mnras, 452, 158

\bibitem[{{Yanny} {et~al}\mbox{.}(2009){Yanny}, {Rockosi}, {Newberg}, {Knapp},
  {Adelman-McCarthy}, {Alcorn}, {Allam}, {Allende Prieto}, {An}, {Anderson},
  {Anderson}, {Bailer-Jones}, {Bastian}, {Beers}, {Bell}, {Belokurov},
  {Bizyaev}, {Blythe}, {Bochanski}, {Boroski}, {Brinchmann}, {Brinkmann},
  {Brewington}, {Carey}, {Cudworth}, {Evans}, {Evans}, {Gates}, {G{\"a}nsicke},
  {Gillespie}, {Gilmore}, {Nebot Gomez-Moran}, {Grebel}, {Greenwell}, {Gunn},
  {Jordan}, {Jordan}, {Harding}, {Harris}, {Hendry}, {Holder}, {Ivans},
  {Ivezi{\v c}}, {Jester}, {Johnson}, {Kent}, {Kleinman}, {Kniazev},
  {Krzesinski}, {Kron}, {Kuropatkin}, {Lebedeva}, {Lee}, {French Leger},
  {L{\'e}pine}, {Levine}, {Lin}, {Long}, {Loomis}, {Lupton}, {Malanushenko},
  {Malanushenko}, {Margon}, {Martinez-Delgado}, {McGehee}, {Monet}, {Morrison},
  {Munn}, {Neilsen}, {Nitta}, {Norris}, {Oravetz}, {Owen}, {Padmanabhan},
  {Pan}, {Peterson}, {Pier}, {Platson}, {Re Fiorentin}, {Richards}, {Rix},
  {Schlegel}, {Schneider}, {Schreiber}, {Schwope}, {Sibley}, {Simmons},
  {Snedden}, {Allyn Smith}, {Stark}, {Stauffer}, {Steinmetz}, {Stoughton},
  {SubbaRao}, {Szalay}, {Szkody}, {Thakar}, {Sivarani}, {Tucker}, {Uomoto},
  {Vanden Berk}, {Vidrih}, {Wadadekar}, {Watters}, {Wilhelm}, {Wyse}, {Yarger},
  \& {Zucker}}]{Yanny+2009}
{Yanny} B. {et~al.}, 2009, \aj, 137, 4377

\bibitem[{{Yuan} {et~al}\mbox{.}(2015){Yuan}, {Liu}, {Huo}, {Xiang}, {Huang},
  {Chen}, {Zhang}, {Sun}, {Wang}, {Zhang}, {Zhao}, {Luo}, {Shi}, {Li}, {Yuan},
  {Dong}, {Li}, {Hou}, \& {Zhang}}]{Yuan+2015a}
{Yuan} H.-B. {et~al.}, 2015, \mnras, 448, 855

\bibitem[{{Yuan}, {Liu} \& {Xiang}(2013){Yuan}, {Liu}, \& {Xiang}}]{Yuan+2013}
{Yuan} H.~B., {Liu} X.~W., {Xiang} M.~S., 2013, \mnras, 430, 2188

\bibitem[{{Zhang} {et~al}\mbox{.}(2014){Zhang}, {Liu}, {Yuan}, {Zhao}, {Yao},
  {Zhang}, {Xiang}, \& {Huang}}]{Zhang+2014}
{Zhang} H.-H., {Liu} X.-W., {Yuan} H.-B., {Zhao} H.-B., {Yao} J.-S., {Zhang}
  H.-W., {Xiang} M.-S., {Huang} Y., 2014, Research in Astronomy and
  Astrophysics, 14, 456

\bibitem[{{Zhang} {et~al}\mbox{.}(2005){Zhang}, {Wu}, {Luo}, \&
  {Zhao}}]{Zhang+2005}
{Zhang} J.~N., {Wu} F.~C., {Luo} A.~L., {Zhao} Y.~H., 2005, Acta Astronomica
  Sinica, 46, 406

\bibitem[{{Zhao} {et~al}\mbox{.}(2012){Zhao}, {Zhao}, {Chu}, {Jing}, \&
  {Deng}}]{Zhao+2012}
{Zhao} G., {Zhao} Y.-H., {Chu} Y.-Q., {Jing} Y.-P., {Deng} L.-C., 2012,
  Research in Astronomy and Astrophysics, 12, 723

\bibitem[{{Zucker} {et~al}\mbox{.}(2012){Zucker}, {de Silva}, {Freeman},
  {Bland-Hawthorn}, \& {Hermes Team}}]{Zucker2012}
{Zucker} D.~B., {de Silva} G., {Freeman} K., {Bland-Hawthorn} J., {Hermes
  Team}, 2012, in Astronomical Society of the Pacific Conference Series, Vol.
  458, Galactic Archaeology: Near-Field Cosmology and the Formation of the
  Milky Way, {Aoki} W., {Ishigaki} M., {Suda} T., {Tsujimoto} T., {Arimoto} N.,
  eds., p. 421

\bibitem[{{Zwitter} {et~al}\mbox{.}(2008){Zwitter}, {Siebert}, {Munari},
  {Freeman}, {Siviero}, {Watson}, {Fulbright}, {Wyse}, {Campbell}, {Seabroke},
  {Williams}, {Steinmetz}, {Bienaym{\'e}}, {Gilmore}, {Grebel}, {Helmi},
  {Navarro}, {Anguiano}, {Boeche}, {Burton}, {Cass}, {Dawe}, {Fiegert},
  {Hartley}, {Russell}, {Veltz}, {Bailin}, {Binney}, {Bland-Hawthorn}, {Brown},
  {Dehnen}, {Evans}, {Re Fiorentin}, {Fiorucci}, {Gerhard}, {Gibson}, {Kelz},
  {Kujken}, {Matijevi{\v c}}, {Minchev}, {Parker}, {Pe{\~n}arrubia}, {Quillen},
  {Read}, {Reid}, {Roeser}, {Ruchti}, {Scholz}, {Smith}, {Sordo}, {Tolstoi},
  {Tomasella}, {Vidrih}, \& {Wylie-de Boer}}]{Zwitter+2008}
{Zwitter} T. {et~al.}, 2008, \aj, 136, 421

\end{thebibliography}

\label{lastpage}

\end{document}